&amslplain
\documentstyle[12pt,amstex,righttag]{article}
\newcommand{\beq}{\begin{equation}}
\newcommand{\eeq}{\end{equation}}
\newcommand{\beqa}{\begin{eqnarray}}
\newcommand{\eeqa}{\end{eqnarray}}
\newcommand{\ba}{\begin{array}}
\newcommand{\ea}{\end{array}}
\newcommand{\CR}{\nonumber \\}
\newcommand{\pa}{\partial}
\newcommand{\A}{\alpha}
\newcommand{\B}{\beta}

\newcommand{\G}{\gamma}

\newcommand{\s}{\sigma}
\newcommand{\bra}{\langle}
\newcommand{\ket}{\rangle}

\newcommand{\half}{{1\over 2}}

\def\vec#1{\mbox{\boldmath $#1$}}

\def\/{\over}
\newcommand{\bo}[1]{\boldsymbol{#1}}    
\newcommand{\orbif}[1]{\bo{C}^3/\bo{Z}_{#1}}
\newcommand{\pos}{\operatorname{pos}}
\newcommand{\sol}{\hat{U}}
\newcommand{\blowup}[1]{\operatorname{Bl}_{\,\bo{\nu}_0}
(\orbif{m})}                               
\newcommand{\Ptwo}{\text{\bf P}^2}
\newcommand{\Hom}{H\!om}
\newcommand{\End}{End}

\begin{document}
\begin{titlepage}
\begin{flushright}
{\tt hep-th/0009072} \\
UTHEP-432 \\
September, 2000
\end{flushright}
\vspace{0.5cm}
\begin{center}
{\Large \bf 
Closed Sub-Monodromy Problems, Local Mirror Symmetry 
and Branes on Orbifolds\par}
\lineskip .75em
\vskip2.5cm
{\large Kenji Mohri, Yoko Onjo and Sung-Kil Yang}
\vskip 1.5em
{\large\it Institute of Physics, University of Tsukuba \\
\vspace{1.5mm}
Ibaraki 305-8571, Japan}
\end{center}
\vskip3cm
\begin{abstract}
We study D-branes wrapping an exceptional four-cycle $\text{\bf P}(1,a,b)$
 in a 
blown-up $\vec{C}^3/\vec{Z}_m$ non-compact Calabi--Yau threefold with 
$(m;a,b)=(3;1,1), (4;1,2)$ and $(6;2,3)$. In applying the method of local 
mirror symmetry we find that the Picard--Fuchs equations for the local mirror 
periods in the $\vec{Z}_{3,4,6}$ orbifolds take the same form as the ones in 
the local $E_{6,7,8}$ del Pezzo models, respectively. It is observed, however,
that the orbifold models and the del Pezzo models possess different physical 
properties because the background NS B-field is turned on in the case of 
$\vec{Z}_{3,4,6}$ orbifolds. This is shown by analyzing the periods and their
monodromies in full detail with the help of Meijer G-functions. We use the
results to discuss D-brane configurations on $\text{\bf P}(1,a,b)$ as well as 
on del Pezzo surfaces. We also discuss the number theoretic aspect of local
mirror symmetry and observe that the exponent which governs the exponential 
growth of the Gromov--Witten invariants is determined by the special value of
the Dirichlet $L$-function.
\end{abstract}
\end{titlepage}

\tableofcontents

\baselineskip=0.7cm

\vskip10mm

\section{Introduction}
\label{sec:Introduction}

Type II string compactification has aroused a great deal of interest in
D-branes on Calabi--Yau space \cite{Douglas}. 
Among recent works \cite{BDLR}--\cite{DiaconescuDouglas},
Diaconescu and Gomis studied the blown-up 
$\vec{C}^3/\vec{Z}_3$ model \cite{DiacGom} 
and found an interesting
correspondence between $\vec{Z}_3$ fractional branes at the orbifold point and
wrapped BPS D-branes on an exceptional $\text{\bf P}^2$ cycle. 
The spectrum of BPS 
D-branes is studied further 
in \cite{DouglasFiolRomelsberger,DouglasFiolRomelsb}.
As demonstrated in these papers,
blown-up orbifolds as models of Calabi--Yau threefolds are worth of being 
considered since they admit an exact description in terms of CFT
at the orbifold point in the K\"ahler moduli space which parameterizes the size
of exceptional four-cycles,
while the large radius behavior of D-branes wrapped on exceptional cycles
can be analyzed by invoking local mirror symmetry \cite{ChKlYaZa}. 
Our purpose in this paper
is to generalize \cite{DiacGom} and consider a blown-up $\vec{C}^3/\vec{Z}_m$ 
model with $m=3,4,6$ in which there exists an exceptional divisor 
$\text{\bf P}^2, \text{\bf P}(1,1,2)$ and
$\text{\bf P}(1,2,3)$, respectively.

The paper is organized and summarized as follows:

In section 2, we start with reviewing a toric description of the blown-ups
of orbifolds $\vec{C}^3/\vec{Z}_m$, and introduce GKZ equations 
for the purpose of applying local mirror symmetry. 
It is seen that our $\vec{Z}_{3,4,6}$ orbifold models are three particular 
examples of non-compact Calabi--Yau threefolds 
${\cal O}_{\text{\bf P}(1,a,b)}(-m)$ 
with $m=1+a+b$. Upon formulating sub-monodromy problems
based on the GKZ equations, we observe that the $\vec{Z}_{3,4,6}$ orbifold 
models and the local $E_{6,7,8}$ del Pezzo models share the Picard--Fuchs 
equations
which are closely related to the $E_{6,7,8}$ elliptic singularities.

In section 3, the detailed analysis of the solutions to the Picard--Fuchs 
equations is presented. Especially we employ Meijer G-functions in constructing
solutions as they provide the natural basis to determine the mirror map.
Moreover, remarkable relations between the special values of G-functions
and zeta functions are observed. This point is considered
further in the next section.

In section 4, the mirror maps for the orbifold models and the local del Pezzo 
models are obtained. It is seen clearly that the difference between the two
models lies in the dependence on the background NS B-field; the B-field is
non-vanishing for the orbifold models, whereas $B=0$ for the del Pezzo models.
We then describe the computation of Gromov--Witten invariants of the models,
putting emphasis on the relation to modular functions.
We also discuss our observation which reveals some arithmetic properties of
local mirror symmetry in view of the relation between the special values of 
zeta functions and the Mahler measure in number theory.

In section 5, we express the BPS central charge in terms of the period
integrals. It is shown that in the large radius limit the same form of the
central charge (up to world-sheet instanton corrections) is derived by the
geometrical consideration of relevant four-cycles embedded 
in Calabi--Yau space.
Combining this observation with the results obtained in the previous sections,
we discuss D-brane configurations on $E_{6,7,8}$ del Pezzo surfaces 
and $\text{\bf P}(1,a,b)$. 

In Appendix A, we review exceptional bundles on $\text{\bf P}^2$
which are relevant to the $\vec{Z}_3$ orbifold model.


\section{Picard--Fuchs equations for local Calabi--Yau}
\renewcommand{\theequation}{2.\arabic{equation}}\setcounter{equation}{0}

\subsection{Toric geometry of orbifolds}

Let us consider the non-compact Calabi--Yau orbifold model
$\orbif{m}$, where the action of the  cyclic group $\bo{Z}_m$
on the coordinates of $\bo{C}^3$ is defined by 
\begin{equation} 
(x_1,x_2,x_3)\rightarrow (\omega x_1,\omega^a x_2,\omega^b x_3).
\end{equation}
Here  $\omega=e^{2\pi i/m}$ is a primitive $m$th root of unity
and the two positive integers $(a,b)$ must satisfy 
the Calabi--Yau condition $1+a+b=m$. 

Toric geometry \cite{Oda,Fulton}
is a powerful tool to describe the blow-ups of 
the orbifold $\orbif{m}$.
Let $N$ be the rank three lattice the generators of which we denote by
$\{\bo{e}_1,\bo{e}_2,\bo{e}_3\}$ and $M=N^{*}$ the dual lattice.
Then $\orbif{m}$ itself admits a toric description by  
the fan ${\cal F}$ defined by a unique maximal cone in $N_{\bo{R}}$:
$
\sigma=\operatorname{pos}\, \{\bo{\nu}_1,\bo{\nu}_2,\bo{\nu}_3\},
$
where 
$\bo{\nu}_1=-a\,\bo{e}_1-b\,\bo{e}_2+\bo{e}_3$,
$\bo{\nu}_2=\bo{e}_1+\bo{e}_3$, 
$\bo{\nu}_3=\bo{e}_2+\bo{e}_3$, and
$\operatorname{pos}\, \{\bo{v}_i\,|\, i\in I\}
:=\oplus_{i\in I}\bo{R}_{\geq 0}\,\bo{v}_i$
means the convex polyhedral cone defined by the 
positive hull of the vectors inside the braces.
The dual cone $\sigma^{*}$ is the cone in $M_{\bo{R}}$
defined by
$\{\, \bo{w}\in M_{\bo{R}}\,|\,\bra\bo{w},\bo{\nu}_{1,2,3}\ket \geq 0\}$.
It can be seen that the ring of the $\bo{Z}_m$-invariant monomials, 
that is the affine coordinate ring of the orbifold 
$\orbif{m}$, is isomorphic to the (additive) semi-group of
the lattice points of the dual cone $M\cap \sigma^{*}$ by
\begin{equation}
M\cap \sigma^{*}\ni \bo{w}
\rightarrow 
x_1^{\bra \bo{w},\bo{\nu}_1\ket}
x_2^{\bra \bo{w},\bo{\nu}_2\ket}
x_3^{\bra \bo{w},\bo{\nu}_3\ket}.
\end{equation} 

Crepant blow-ups of a variety are those which preserve
its canonical line bundle; in particular, 
a crepant blow-up of a Calabi--Yau variety respects 
the Calabi--Yau condition, as it is equivalent to
the triviality of the canonical line bundle.
For the case of our orbifold $\orbif{m}$, 
it is known that there is a one-to-one correspondence
between the crepant divisors and the set of the lattice points  
\begin{equation}
\{\, \bo{\nu}\in \sigma\cap N\,|\, \bra \bo{e}_3^{*}, \bo{\nu}\ket =1\,\},
\label{crepant}
\end{equation}
which are incorporated in the refinement of the fan ${\cal F}$
under the corresponding blow-up.

Let us consider the crepant (partial) blow-up
$\blowup{m} \rightarrow \orbif{m}$
defined by the subdivision of the cone $\sigma$ 
by the vector $\bo{\nu}_0:=\bo{e}_3$ 
which is an element of (\ref{crepant}).
In the process of the blow-up, the origin $(0,0,0)$ is blown-up
to the exceptional divisor $\text{\bf P}(1,a,b)$, 
and the resulting Calabi--Yau variety 
$\blowup{m}$ is identified with
the canonical line bundle 
(in the orbifold sense) of it, that is, we have
\begin{equation}
\blowup{m}
\cong K_{\text{\bf P}(1,a,b)}
={\cal O}_{\text{\bf P}(1,a,b)}(-m).
\label{orbifoldlinebundle}
\end{equation}
The fan of the blown-up orbifold $\blowup{m}$, which 
we denote by $\widetilde{\cal F}$, 
is defined by the collection of the following three maximal cones:
\begin{equation*}
\sigma_1=\pos\, \{\bo{\nu}_0,\bo{\nu}_2,\bo{\nu}_3\},\quad
\sigma_2 =\pos\, \{\bo{\nu}_0,\bo{\nu}_1,\bo{\nu}_3\},\quad
\sigma_3 =\pos\, \{\bo{\nu}_0,\bo{\nu}_1,\bo{\nu}_2\}.
\end{equation*}
These maximal cones define the affine open covering 
$\blowup{m}
=\bigcup_{i=1}^{3}\, U_{\sigma_i}$, where 
$U_{\sigma_1}\cong \bo{C}^3$
is a smooth patch, however the remaining two 
$U_{\sigma_2}\cong \orbif{a}$,
$U_{\sigma_3}\cong \orbif{b}$
have orbifold singularities in general.
The exceptional divisor $S:=\text{\bf P}(1,a,b)$ is the one 
associated with the 1-cone $\bo{R}_{\geq 0}\,\bo{\nu}_0$ 
in $\widetilde{\cal F}$, the toric description of which is given as follows:
Let $\pi:N\to {\bar N}=N/\bo{Z}\, \bo{e}_3$ the quotient lattice and
the canonical projection. 
Then the two dimensional complete fan ${\bar {\cal F}}$
defined by the collection of the maximal cones
$\pi(\sigma_1)$,
$\pi(\sigma_2)$ and
$\pi(\sigma_3)$
in ${\bar N}_{\bo{R}}$ produces $\text{\bf P}(1,a,b)$ as the associated
toric twofold. It is seen that $\text{\bf P}(1,a,b)$  has $\bo{Z}_a$ 
and $\bo{Z}_b$ orbifold singular points.
We can compute its triple intersection in the blown-up orbifold:
\begin{equation}
S \cdot S \cdot S = c_1(S)\cdot c_1(S)=\frac{m^2}{ab}.
\label{triple-orb}
\end{equation}   
The convex polyhedron in ${\bar N}_{\bo{R}}$
defined by the convex hull of the three points: 
$\pi(\bo{\nu}_1)$, $\pi(\bo{\nu}_2)$, $\pi(\bo{\nu}_3)$,
becomes a {\em reflexive polyhedron}
only in the three cases:
$\{ a, b \}$ $=$ $\{ 1, 1 \}$, $\{ 1, 2 \}$, $\{ 2, 3 \}$,
when the exceptional divisor $\text{\bf P}(1,a,b)$ has 
as its anti-canonical divisor an elliptic curves of the type $E_{6,7,8}$
respectively. 
The connection between non-compact orbifolds 
and elliptic curves in these distinguished models
will become important when we solve  the Picard--Fuchs 
equations of them below.

The introduction of the homogeneous coordinates $(x_0,x_1,x_2,x_3)$
greatly simplifies the construction of the blown-up orbifold
$\blowup{m}$, where each coordinate $x_i$ corresponds 
to the primitive generator $\bo{\nu}_i$ 
and the linear relation between them
\begin{equation}
-m\,\bo{\nu}_0+\bo{\nu}_1+a\,\bo{\nu}_2+b\,\bo{\nu}_3=\text{\bf 0}
\end{equation}  
tells us the U(1) charge assignment for the homogeneous coordinates:
\begin{equation}
(x_0;x_1,x_2,x_3)
\sim 
(\lambda^{-m}\,x_0; \lambda\,x_1, \lambda^{a}\,x_2, \lambda^{b}\,x_3),
\quad \lambda\in \bo{C}^{*},
\end{equation}
where $x_0$ represents the fiber direction of the orbifold line bundle
(\ref{orbifoldlinebundle}), and $(x_1,x_2,x_3)$  the homogeneous 
coordinates of the base twofold $\text{\bf P}(1,a,b)$.
The charge vector 
\begin{equation}
l=(l_i)=(-m;1,a,b),
\label{Mori-orb}
\end{equation}
is called the  Mori vector, from which we can 
write down the Picard--Fuchs equation for the local mirror
periods of the blown-up orbifold $\blowup{m}$.

\subsection{GKZ equations for orbifolds}
\label{P-F for orbifold}

There is a standard procedure to derive the Picard--Fuchs 
equation for the blown-up orbifold $\blowup{m}$ from its 
toric data \cite{Ba,HKTY,AsGrMo,As}, 
which we review briefly here.
First let us define the bare K\"ahler modulus parameter $z$,
which controls the size of the exceptional divisor, by
\begin{equation}
z=\prod_{i=0}^{3}\left(\frac{a_i}{l_i}\right)^{l_i}:=e^{\beta}\,z_0,
\quad e^{\beta}=\prod_{i=0}^{3}\left| l_i^{-l_i}\right|, 
\label{z-beta-z0}
\end{equation}
where $\{a_i\}$ are the coefficients of the monomials
appearing in the defining polynomial of the mirror variety,
and we use either $z$ (normalized) or $z_0$ (unnormalized) 
according to the situation.  
Note that the large radius region corresponds to $|z|\ll 1$,
while the region with $|z|\gg 1$ is called the Landau--Ginzburg 
or orbifold phase. 
Second, given a  general Mori vector $(l_i)$,
the GKZ operator associated with it is 
\begin{equation}
\boxed{\phantom{\text{a}}}_{\,l}
:=\prod_{l_i >0}\left({\pa\/\pa a_i}\right)^{l_i}
     -\prod_{l_i <0}\left({\pa\/\pa a_i}\right)^{-l_i}.
\label{GKZ-general}
\end{equation}
In particular, for our blown-up orbifold, 
the use of (\ref{Mori-orb}) combined with the ansatz for a mirror period
$\varPi(a_i)=f(z)$ leads to the following GKZ equation 
\cite{As}:
\begin{align}
\boxed{\phantom{\text{a}}}_{\,\text{orb}}\,f(z)
&=0,\nonumber \\
\boxed{\phantom{\text{a}}}_{\,\text{orb}}
&=\left\{
\prod_{k_2=0}^{a-1}\big(\varTheta_z-\frac{k_2}{a}\big)
\prod_{k_3=0}^{b-1}\big(\varTheta_z-\frac{k_3}{b}\big)
-z\prod_{k_0=1}^{m-1}\big(\varTheta_z+\frac{k_0}{m}\big)
\right\}\circ\varTheta_z,
\label{GKZ-orb}
\end{align}
where 
$\varTheta_z=zd/dz$ is the logarithmic differential operator as usual.
Let us consider the behavior of the solutions of (\ref{GKZ-orb})
around the large radius limit point $z=0$, where we can rely 
on the classical geometry of the exceptional divisor $\text{\bf P}(1,a,b)$.
Substituting the ansatz 
$f(z)=\sum_{n=0}^{\infty} f_n  z^{n+\rho}$
for a solution of (\ref{GKZ-orb}),
we obtain the indicial equation  for $\rho$:
\begin{equation}
\prod_{k_2=1}^{a-1} \left(\rho-\frac{k_2}{a}\right)
\prod_{k_3=1}^{b-1} \left(\rho-\frac{k_3}{b}\right)\cdot
\rho^3=0.
\end{equation}
The triple zero at $\rho=0$ yields the three solutions 
of the GKZ equation (\ref{GKZ-orb}): the constant solution 1, 
the single- and double-log solutions, which clearly correspond to 
the zero-, two- and four-cycles on the exceptional divisor.

The most efficient way to obtain these solutions would be
the Frobenius method \cite{HKTY}; 
We first make the formal power series
\begin{align}
\sol_0(z,\rho)&=\sum_{n=0}^{\infty}
{\bar A}(n+\rho)(e^{\epsilon\pi i} z_0)^{n+\rho},
\quad \epsilon=
\begin{cases}
0, &m=\text{even},\\
1,&m=\text{odd},
\end{cases}
\\
{\bar A}(n)&=\frac{1}{\varGamma(-mn+1)
\varGamma(n+1)\varGamma(an+1)\varGamma(bn+1)}.
\end{align} 
The three solutions
\footnote[2]{If  $\left\{ {k_2}/{a},{k_3}/{b}\right\}$ $\cap$
$\left\{{k_0}/{m}\right\}$ is not empty,
we can delete the corresponding factors from the left of the GKZ operator
(\ref{GKZ-orb}), to get a operator of lower rank, as we shall do
in (\ref{P-F zE7}) and  (\ref{P-F zE8}).
The three functions  $1, \sol_1(z), \sol_2(z)$ 
obtained by the Frobenius method
are the solutions of this reduced GKZ equation.}  
then are recovered by the expansion 
in the formal variable $\rho$:
\begin{equation*}
\lim_{\rho\to 0}\sol_0(z,\rho)=1,\quad
\lim_{\rho\to 0}\frac{\partial}{\partial\rho}\,\sol_0(z,\rho),\quad
\lim_{\rho\to 0}\frac12 \frac{\partial^2}{\partial\rho^2}\,\sol_0(z,\rho).
\end{equation*}
For completeness, we give the explicit forms 
of the two non-trivial solutions:
\begin{align}
\sol_1(z)&=\log(z_0)
+\sum_{n=1}^{\infty}A_0(n)\,z_0^n, 
\label{single-log}\\
\sol_2(z)&=\frac12\log^2(z_0)
+\sum_{n=1}^{\infty}A_0(n)\,z_0^n\, \log(z_0)
+\sum_{n=1}^{\infty}A_0(n)B(n)\,z_0^n,
\label{double-log}
\end{align}
where
\beqa
A_0(n)
\!&=& \!\frac{\varGamma(mn+1)}{n\varGamma(n+1)\varGamma(an+1)\varGamma(bn+1)},
\label{keisuu-A}\\
B(n)\!&=& \! m\,\varPsi(mn+1)
-\varPsi(n+1)-a\,\varPsi(an+1)-b\,\varPsi(bn+1)-\frac1n,
\label{keisuu-B}
\eeqa
and $\varPsi(x)={d\/dx}\log\varGamma(x)$ is the digamma function.
Note that the single-log solution given in \cite{As} coincides with 
(\ref{single-log}).
\footnote[3]{The factor $(N_l n + 1)!$ in (28) of \cite{As} should read 
$(N_l n - 1)!$.}

On the other hand,
the solutions of the GKZ equation (\ref{GKZ-orb})
associated with fractional $\rho=k_2/a$, $k_3/b$
are {\em unphysical}, 
which must be abandoned because our interest is
only in the BPS D-brane system on the non-compact orbifolds.  

In fact, the use of the {\em Meijer G-functions} (see the next section)
enables us to study systematically the {\em closed sub-monodromies} 
of the three periods $\{1,\sol_1(z),\sol_2(z)\}$  
not only around  the large radius limit point $z=0$, 
but also around the Landau--Ginzburg point $z=\infty$ 
(hence also around the discriminant locus $z=1$).
However, instead of 
treating the general orbifold models rather abstractly,
we will  restrict ourselves below to the three distinguished models, 
because the connection of them with 
the local $E_{6,7,8}$ del Pezzo models is very interesting, 
and that with the $E_{6,7,8}$ tori greatly facilitates 
the exact analysis of the Picard--Fuchs system of the orbifolds.

\subsection{Three distinguished models}

The three distinguished orbifolds 
mentioned in the last paragraph of the preceding subsection are 
$(m;a,b)$ $=$ $(3;1,1)$, $(4;1,2)$ and $(6;2,3)$,
which we call  
$\bo{Z}_3$,
$\bo{Z}_4$ and 
$\bo{Z}_6$ models  for simplicity. 

For these models,
it is possible to factorize an appropriate 
Picard--Fuchs operator of rank three 
${\cal L}_{\text{orb}}$ on the right of the GKZ operator
$\boxed{\phantom{\text{a}}}_{\,\text{orb}}$,
the three solutions of which close under the monodromy actions
and indeed correspond to  the zero, two- and four-cycles 
on the exceptional divisor.
In fact,  
the GKZ operators (\ref{GKZ-orb})  of 
$\bo{Z}_3$,
$\bo{Z}_4$ and 
$\bo{Z}_6$ models
admit respectively the following factorizations:
\beqa
\boxed{\phantom{\text{a}}}_{\text{\,orb}} &=&  
 \left\{\varTheta_z^2
  -z\big(\varTheta_z+{1\/3}\big)
\big(\varTheta_z+{2\/3}\big)\right\}\circ\varTheta_z,\
\label{P-F zE6}\\
\boxed{\phantom{\text{a}}}_{\text{\,orb}} &=& 
\big(\varTheta_z-{1\/2}\big)\circ
\left\{\varTheta_z^2
 -z\big(\varTheta_z+{1\/4}\big)
\big(\varTheta_z+{3\/4}\big)\right\}
\circ\varTheta_z, \
\label{P-F zE7} \\
\boxed{\phantom{\text{a}}}_{\text{\,orb}} &=& 
\big(\varTheta_z-{1\/3}\big)
\big(\varTheta_z-{1\/2}\big)
\big(\varTheta_z-{2\/3}\big)
\circ\left\{\varTheta_z^2
-z\big(\varTheta_z+{1\/6}\big)
\big(\varTheta_z+{5\/6}\big)\right\}
\circ\varTheta_z.\ 
\label{P-F zE8}
\eeqa
Hence we can define the Picard--Fuchs  operator by
\begin{equation}
{\cal L}_{\text{orb}}
={\cal L}_{\text{ell}}\circ \varTheta_z
=\left\{\varTheta_z^2
-z(\varTheta_z+\alpha_1)(\varTheta_z+\alpha_2)\right\}
\circ \varTheta_z,
\end{equation}
where $(\alpha_1,\alpha_2)=$
$(\frac13,\frac23)$, 
$(\frac14,\frac34)$,  
$(\frac16,\frac56)$,
for 
$\bo{Z}_3$, 
$\bo{Z}_4$, 
$\bo{Z}_6$ 
orbifold model respectively, and ${\cal L}_{\text{ell}}$
is the Picard--Fuchs operator of the torus 
which shares the same toric data
(\ref{orbifoldlinebundle}) with the corresponding orbifold, 
but has the different ansatz: $\varPi(a_i)=f(z)/a_0$ for its periods.

\subsection{Picard--Fuchs equations for local del Pezzo models}

In this subsection, we collect the facts about the toric description 
of the three local del Pezzo models and their Picard--Fuchs equations
\cite{KlemmMayrVafa,LeMayWar},
which are closely related to those of the three orbifold models 
described in the previous subsection, for convenience.

$E_{6,7,8}$ del Pezzo surfaces $S_{6,7,8}$ 
can be realized as the hypersurfaces
in weighted projective threefolds:
\begin{alignat}{2}
&E_6: \quad  &  &\text{\bf P}(1,1,1,1)[3],\\
&E_7: \quad &  &\text{\bf P}(1,1,1,2)[4],\\
&E_8: \quad &  &\text{\bf P}(1,1,2,3)[6].
\end{alignat}
If one of them, which we denote by $S_N$, $N=6,7,8$, 
is embedded in a compact Calabi--Yau threefold $X$, 
then the neighborhood of $S_N$ in $X$ is identified with
the canonical line bundle of $S_N$:
$K_{S_N}\cong {\cal O}_{S_N}(-1)$,
where the right hand side is the restriction to the 
hypersurface $S_N$ of the orbifold line bundle 
${\cal O}_{\text{\bf P}(1,1,a,b)}(-1)$
on the weighted projective threespace with 
$(a,b)=$ 
$(1,1)$,
$(1,2)$,
$(2,3)$ 
for $N=6,7,8$ respectively.

The triple intersection of the  $E_{N}$ del Pezzo surface
$S_N$ embedded in a Calabi--Yau threefold $X$ is computed as
\begin{equation}
S_N\cdot S_N \cdot S_N=c_1(S_N)\cdot c_1(S_N)=9-N.
\label{triple-dP}
\end{equation}

We see that the non-compact toric Calabi--Yau fivefold 
associated with the local del Pezzo model is
the rank two orbifold bundle on $\text{\bf P}:=\text{\bf P}(1,1,a,b)$:  
\begin{equation}
{\cal O}_{\text{\bf P}}(-m)\oplus {\cal O}_{\text{\bf P}}(-1).
\label{bundle-delPezzo}
\end{equation}
In fact, this toric data is shared with both the $E_{6,7,8}$ torus 
and the $\bo{Z}_{3,4,6}$ blown-up orbifold model,
because the former can be realized as a complete intersection 
$\text{\bf P}(1,1,a,b)[1,m]$ and the exceptional divisor 
of the latter as a hypersurface $\text{\bf P}(1,1,a,b)[1]$.

A realization of (\ref{bundle-delPezzo}) 
by means of the homogeneous coordinates, 
the first two of which represent the non-compact directions,  becomes
\begin{equation}
(x_{-1},x_0;x_1,x_2,x_3,x_4)\sim
(\lambda^{-1}\, x_{-1},\lambda^{-m}\,x_0;\lambda\, x_1,\lambda\, x_2,
\lambda^{a}\,x_3,\lambda^b\,x_4),
\end{equation}
{}from which we identify the Mori vector as
$l=(-1,-m;1,1,a,b)$, that is,
\begin{alignat}{2}
&E_6:\quad  & l&=(-1,-3;1,1,1,1),\\
&E_7:\quad  & l&=(-1,-4;1,1,1,2),\\
&E_8:\quad  & l&=(-1,-6;1,1,2,3).
\end{alignat}

%
The formula of the GKZ operator for a given Mori vector $l$
(\ref{GKZ-general}) gives the GKZ equation for 
the local del Pezzo models under the ansatz for the periods 
$\varPi(a_i)=f(z)/a_0$: 
\begin{equation}
\boxed{\phantom{\text{a}}}_{\, \text{dP}}
 =\varTheta_z\circ\boxed{\phantom{\text{a}}}_{\, \text{orb}},
\end{equation}
where the $E_{6,7,8}$ del Pezzo models correspond to the
$\bo{Z}_{3,4,6}$ orbifold models respectively.
Note that  the GKZ equations for the $E_{6,7,8}$ torus
and the $\bo{Z}_{3,4,6}$ orbifold model can be obtained 
if we take
$\varPi(a_i)=f(z)/(a_{-1}a_0)$ and  
$\varPi(a_i)=f(z)/a_{-1}$ for the periods respectively. 

To summarize, the relations among the Picard--Fuchs operators
of $\bo{Z}_{3,4,6}$ orbifolds, $E_{6,7,8}$ del Pezzo surfaces 
and $E_{6,7,8}$ tori become   
\begin{equation}
{\cal L}_{\text{dP}}={\cal L}_{\text{orb}}
={\cal L}_{\text{ell}}\circ\varTheta_z
=\left\{\varTheta_z^2
-z(\varTheta_z+\alpha_1)(\varTheta_z+\alpha_2)\right\}
\circ\varTheta_z,
\label{LdP-Lorb-Lell}
\end{equation}
where $(\A_1,\A_2)$ takes
\beqa
 (\A_1,\A_2)=\left({1\/3},\, {2\/3}\right),\quad
             \left({1\/4},\, {3\/4}\right),\quad
             \left({1\/6},\, {5\/6}\right)
\label{A1A2}
\eeqa
for the $\vec{Z}_{3,4,6}$ (or $E_{6,7,8}$) models respectively.


\section{Solutions of Picard--Fuchs equations}
\label{sec: Solutions of Picard--Fuchs equation and monodromy}
\renewcommand{\theequation}{3.\arabic{equation}}\setcounter{equation}{0}

The Picard--Fuchs equations ${\cal L}_{\text{ell}}\circ \varTheta_z\varPi=0$
 have already appeared in the literature 
\cite{DiacGom,ChKlYaZa,AsGrMo,KlemmMayrVafa,LeMayWar,KlemmZaslow,MOY} 
in the context of local mirror symmetry and D-brane physics.
Since the Picard--Fuchs operator has the factorized form 
${\cal L}_{\text{ell}} \circ \varTheta_z$ one may obtain the solution 
by performing the logarithmic integral of the torus periods $\varpi(z)$ 
which obey ${\cal L}_{\text{ell}}\varpi(z)=0$.
See \cite{MOY} for a recent thorough treatment along this line 
in the case of the del Pezzo models.
It has been recognized, however, that the method of Meijer G-functions is 
more systematic in dealing with the generalized hypergeometric equation
\cite{KlemmZaslow,GrLa,Lazaroiu}.
In particular, the analytic continuation of periods between a patch $|z|<1$ 
(the large radius region) and a patch $|z|>1$ 
(the orbifold/Landau-Ginzburg region) can be performed unambiguously.
It also turns out that Meijer G-functions provide a suitable set of
fundamental solutions in constructing a mirror map as will be observed 
in section \ref{sec: Mirror map}. Thus we think of it worth presenting the 
details of the analysis with the use of Meijer G-functions.

Meijer G-functions are defined by \cite{Bate}
\beqa 
&&\!\!\!
\text{G}_{r+r',s+s'}^{s,r}
\left(\left.\ba{cccccc}
      \rho_1&\cdots&\rho_r&\rho_{r+1}&\cdots&\rho_{r+r'}\\
      \s_1  &\cdots&\s_s  &\s_{s+1}  &\cdots&\s_{s+s'}  \ea\right| z
\right)\CR
&&
 =\int_{\G}{ ds\/2\pi i} 
  {\varGamma(\s_1-s)\cdots\varGamma(\s_s-s)\varGamma(1-\rho_1+s)\cdots
   \varGamma(1-\rho_r+s)\/
   \varGamma(\rho_{r+1}-s)\cdots\varGamma(\rho_{r+r'}-s)
   \varGamma(1-\s_{s+1}+s)   \cdots\varGamma(1-\s_{s+s'}+s)}  z^s,
\label{Meijer-Gfct def}
\eeqa
where the integration path $\G$ 
runs from $-i\infty$ to $+i\infty$ so as to separate the poles at 
$s=\s_i+n$ from those at $s=-n-1+\rho_i$ with $n$ being the non-negative 
integers.
They satisfy the linear differential equation
\beqa
\left\{
  \prod_{i=1}^{s+s'}(\varTheta_z-\s_i)
     -(-1)^{\mu}z\prod_{j=1}^{r+r'}(\varTheta_z-\rho_j+1)
\right\}\text{G}= 0     
\label{G-fct's diff eq},
\eeqa
where $\mu = r'- s\pmod{2}$.

\begin{figure}[h]
\begin{center}
\input epsf
\hspace{1cm}
\epsfbox{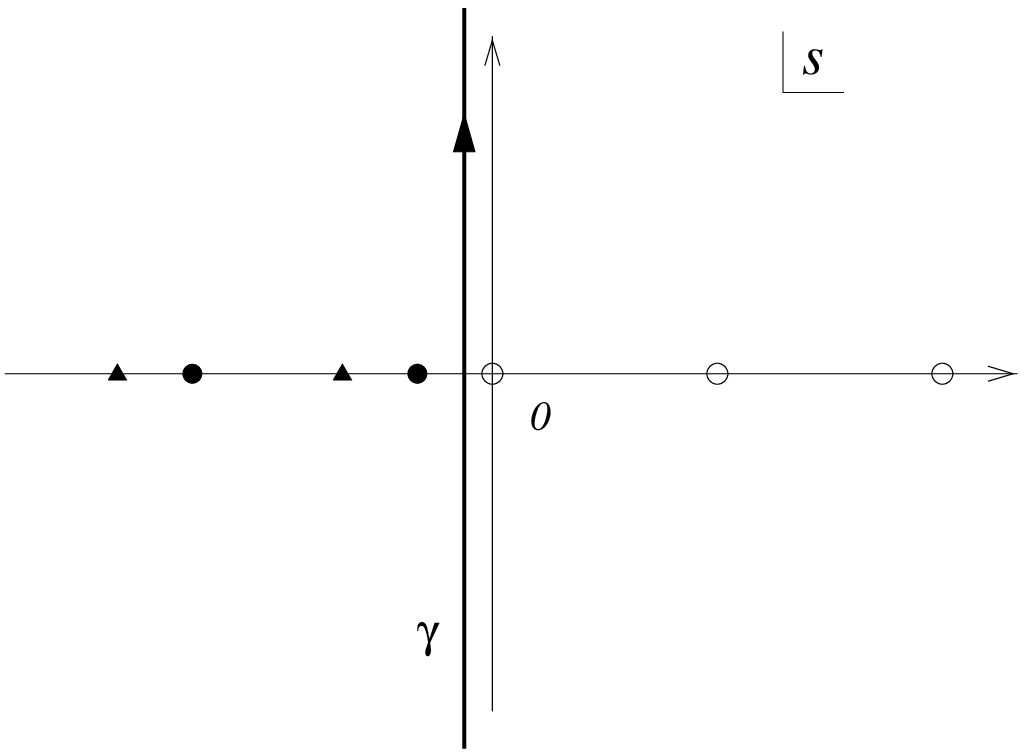}
\caption{The integration path $\gamma$ for $\protect\text{G}_{3,3}^{2,2}$ 
and $\protect\text{G}_{3,3}^{3,2}$.}
\label{contour of G}
\end{center}
\end{figure}

Let us set $r+r'=s+s'=3$ and  $\rho_1=\A_1,\ \rho_2=\A_2,\ \rho_3=1,\ \s_i=0$, 
then (\ref{G-fct's diff eq}) is reduced to our Picard--Fuchs equations with
(\ref{LdP-Lorb-Lell}) which have the regular singular 
points at $z=0,1$ and $\infty$.
It is known that a fundamental system of solutions around $z=0$ 
as well as $z=\infty$ is given by Meijer G-functions \cite{Bate}.
For these regions, thus, solutions to 
${\cal L}_{\text{ell}}\circ \varTheta_z\varPi=0$ are derived from 
Meijer G-functions
\beqa
\text{G}_{3,3}^{s,r}
\left(\left.
        \ba{ccc} \A_1& \A_2&1\\0&0&0\ea
      \right| (-1)^{\mu}z 
\right).
\eeqa
As a fundamental system of solutions 
we take $ (1,U_1(z),U_2(z))$ where
\beqa 
U_1(z)&=& -{\sin{\pi\A_1}\/\pi }
          \,\text{G}_{3,3}^{2,2}
              \left(\left.\ba{ccc}  \A_1&  \A_2& 1 \\
                                     0  &   0  & 0 \ea
                    \right|-z \right)                   \CR
      &=& -{\sin{\pi\A_1}\/2\pi^2 i}
            \int_{\G}{ ds
              {\varGamma(\A_1+s)\varGamma(\A_2+s)\varGamma(-s)^2\/
               \varGamma(1-s)\varGamma(1+s)}(-z)^s} ,   \\
U_2(z)&=& -{\sin{\pi\A_1}\/\pi }
          \,\text{G}_{3,3}^{3,2}
              \left(\left.\ba{ccc}  \A_1&  \A_2& 1 \\
                                     0  &   0  & 0 \ea
                    \right| z \right)                   \CR
      &=& -{\sin{\pi\A_1}\/2\pi^2 i}
            \int_{\G}{ ds
              {\varGamma(\A_1+s)\varGamma(\A_2+s)\varGamma(-s)^3\/
               \varGamma(1-s)}z^s }.
\label{U12 def}
\eeqa
Here a normalization factor $-{\sin{\pi\A_1}/\pi}$, 
which equals $-{1/\varGamma(\A_1)\varGamma(\A_2)}$, has been introduced 
for convenience.
The path $\G$ is depicted in  Fig.\ref{contour of G}.
Examining the asymptotic behavior of integrands as $s\rightarrow \pm i\infty$
with the aid of Stirling's formula, it is shown that the integrals converge 
if $|\arg(-z)|<\pi $ for $U_1(z)$ and $|\arg(z)|< 2\pi $ for $U_2(z)$.
In the following we choose a branch so that
\beqa \log(-z)=\log(z) + i\pi. \eeqa
%

\subsection{Solutions at $z=0$ }
\label{Solutions at $z=0$}

When $|z| < 1$, we can close the contour $\G$ to the right 
and evaluate the integrals as a sum over the residues 
of poles at $s = 0,1,2,\ldots$.
As a result we obtain 
\beqa
U_1(z) &=&    \log\left({-z\/e^{\beta}}\right) 
            +\sum_{n=1}^{\infty}A(n)z^n, 
 \label{U1 z0}\\
U_2(z) &=& \ -\frac12 \log^2\big(\frac{z}{e^{\beta}}\big)
            -\sum_{n=1}^{\infty}A(n) z^n \log\big(\frac{z}{e^{\beta}}\big)
             -\sum_{n=1}^{\infty} A(n) B(n) z^n
             -\xi,
 \label{U2 z0}
\eeqa
where
\beqa
   A(n) &=& {(\alpha_1)_n(\alpha_2)_n\/(n!)^2 n},\CR
   B(n) &=&   \sum_{k=0}^{n-1}\big(\frac{1}{k+\alpha_1}+\frac{1}{k+\alpha_2}
             -\frac{2}{k+1}\big)
             -\frac{1}{n},
\label{B(n)}
\eeqa
and $(\A)_n=\varGamma(\A+n)/\varGamma(\A)$.
Here two constants $\beta$ and $\xi$ are given by
\beqa  
 \beta  &=& -\varPsi(\A_1)-\varPsi(\A_2)+2\varPsi(1),\CR
   \xi  &=&   {1\/2}
           \left(\varPsi'(\A_1) +\varPsi'(\A_2) +2\varPsi'(1)\right).
\eeqa
{}From the special values of $\varPsi(x)$ one can check that 
$e^{\beta}= \prod_i| l_i^{-l_i}|\ \text{(as defined in (\ref{z-beta-z0}))}
= \{27, 64, 432\}$ and $\xi = {\pi ^2/6}\times\{5,7,13\}$ for $(\A_1,\A_2)$ 
given in (\ref{A1A2}). 
For later use, we note the relation 
\beqa 
  U_2(z) = -\half U_1^2(z) + \pi iU_1(z) + {\pi^2\/2} 
           -\xi + \text{O}(z). 
\label{U1U2}
\eeqa

Under $z\rightarrow e^{2\pi i}z$, the monodromy matrix acting on the basis 
\beqa
\varPi_U=\left(1,{U_1(z)\/2\pi i},-{U_2(z)\/(2\pi i)^2}\right)
\eeqa
is obtained as
\beqa M_0 = \left(\ba{ccc}
           1 & 0 & 0 \\
           1 & 1 & 0 \\
           0 & 1 & 1
                 \ea\right) 
\label{M0-monodromy}
\eeqa 
irrespective of the models.

\subsection{Solutions at $z = \infty$}
\label{subsec:Solutions at z=infty}

For $|z| > 1$ the contour $\G$ can be closed to the left.
Then, summing over the residues of poles at $s = -\A_i-n$ 
with non-negative integers $n$ 
we have power series expansions which are expressed in terms of 
generalized hypergeometric functions 
\beqa
\left(\ba{c}
   U_1^{\infty}(\zeta)\\U_2^{\infty}(\zeta)
\ea\right)
 =-\left(\ba{cc} Y_{11}&\ Y_{12}\\ Y_{21}&\ Y_{22}\ea\right)
   \left(\ba{c}
     \zeta^{\A_1}\,
      {}_3\text{F}_2\left(\A_1,\A_1,\A_1;1+\A_1,2\A_1;\zeta\right)\\
     \zeta^{\A_2}\,
      {}_3\text{F}_2\left(\A_2,\A_2,\A_2;1+\A_2,2\A_2;\zeta\right) 
   \ea\right) ,
\eeqa
where $\zeta=1/z$ and
\begin{alignat}{2}
 Y_{11}&=&\ {e^{-i\pi\A_1}
            \varGamma(\A_2-\A_1)\/\A_1\varGamma(\A_2)^2} ,            
\quad 
 Y_{12}&=  Y_{11}(\A_1\leftrightarrow\A_2),
\CR
 Y_{21}&=&\ {\varGamma(\A_1)
            \varGamma(\A_2-\A_1)\/\A_1\varGamma(\A_2)} ,            
\quad 
 Y_{22}&=  Y_{21}(\A_1\leftrightarrow\A_2) .\label{Y}
\end{alignat}

It is easy to see how these solutions are related to Meijer G-functions.
Upon a change of variable $z=1/\zeta$ 
(\ref{LdP-Lorb-Lell}) takes again the Meijer form
\beqa 
\left\{
(\varTheta_{\zeta}-\A_1)
(\varTheta_{\zeta}-\A_2)\varTheta_{\zeta}
                  -\zeta\varTheta_{\zeta}^3
\right\}f=0
\eeqa
whose solutions are given by 
\beqa 
\text{G}_{3,3}^{s,r}
\left({\left.
\ba{ccc}  1 &  1 & 1 \\
        \A_1&\A_2& 0 \ea
\right| (-1)^{s+r+1}\zeta}\right).
\eeqa
Setting $(s,r)=(2,2)$ and $(3,2)$ we find
\begin{align} 
 U_1^\infty (\zeta)&=
-{\sin{\pi \A_1}\/ \pi}\,\text{G}_{3,3}^{2,2}
\left({\left.
\ba{ccc}  1 &  1 & 1 \\
        \A_1&\A_2& 0 \ea
\right| -\zeta}\right),    \nonumber \\
 U_2^\infty (\zeta)&=
-{\sin{\pi \A_1}\/ \pi}\,\text{G}_{3,3}^{3,2}
\left({\left.
\ba{ccc}  1 &  1 & 1 \\
        \A_1&\A_2& 0 \ea
\right| \zeta}\right).
\label{UinfG}
\end{align}

The monodromy matrix at $z=\infty$ is now evaluated to be
\beqa
M_{\infty} 
 = \left(\ba{ccc}
          1  &        0  &        0  \\
          0  & 1-\lambda & -\lambda  \\
          0  &        1  &        1
   \ea\right) ,
\label{M_infty}
\eeqa
where $\lambda=4\sin^2 \pi\A_1=3,2,1$ and $(M_{\infty})^m=I$ 
with $m=3,4,6$ for $(\A_1,\A_2)$ in (\ref{A1A2}).
Thus $\vec{Z}_m$ quantum symmetries are realized at the $z=\infty$ 
orbifold/Landau--Ginzburg points.

Now that monodromies at $z=0$ and $z=\infty$ have been determined, 
one may infer the monodromy matrix $M_1$ at $z=1$ from the relation 
$M_1M_0=M_{\infty}$, 
\beqa 
M_1 &=& M_{\infty}M_0^{-1} 
    = \left(\ba{rrr}
           1  &  0  &       0         \\
          -1  &  1  & -\lambda        \\
           0  &  0  &       1
        \ea\right). 
\label{M_1}
\eeqa
In the next section, we confirm this by explicitly constructing solutions 
at $z=1$.

\subsection{Solutions at $z=1$}

It contract to the previous cases, solutions 
of ${\cal L}_{\text{ell}}\circ\varTheta_z\varPi=0$  around $z=1$ 
cannot be expressed in the form of Meijer G-functions.
In fact, the Picard--Fuchs operators (\ref{LdP-Lorb-Lell}) do not 
take the Meijer form for the variable $u=1-z$.
Thus, in subsection \ref{Picard--Fuchs equation around z=1}, 
we first solve the differential equation recursively, 
and then, in subsection \ref{torus solution}, we give a method 
to construct solutions by the logarithmic integral 
of corresponding torus periods which are given by Meijer G-functions.

\subsubsection{Solutions from the recursion relation}
\label{Picard--Fuchs equation around z=1}

Making a change of variable $u=1-z$, 
we rewrite the Picard--Fuchs equations as
\beqa 
\left\{\varTheta_u^3+{u+2\/u-1}\varTheta_u^2
    +{\A_1\A_2u^2-\A_1\A_2u+1\/(u-1)^2}\varTheta_u\right\}
      \varPi=0.
\label{P-Fz1}
\eeqa
If we set $\varPi=\sum_{n=0}^{\infty} a_nu^{n+\rho}$, 
the indicial equation reads $\rho(\rho-1)^2=0$. 
Thus we have a set of solutions $(1,V_1(u),V_2(u))$,
\beqa 
V_1(u) &=& \sum_{n=0}^{\infty} a_nu^{n+1},\quad a_0=1, 
\label{omega1}\\
V_2(u) &=& V_1(u)\log{u} + \sum_{n=1}^{\infty} b_nu^{n+1}, 
\label{omega2}
\eeqa
where the coefficients $a_n$ and $b_n$ can be determined recursively.

The recursion relations for the coefficients $a_n$ in $V_1$ are
\beqa
a_1={1\/2}(1+\A_1\A_2),\phantom{\{-2(m-1)^2+(m-1)-\A_1\A_2\}}&&\\
m(m+\A_1)(m+\A_2)a_{m-1}+(m+1)\{-2(m+1)^2+(m+1)-\A_1\A_2\}a_{m}&&\CR
+(m+2)(m+1)^2a_{m+1}=0, \qquad  \hbox{for $m\ge 1$}.\qquad&&
\eeqa

The recursion relations obeyed by $b_n$ in $V_2$ are 
\beqa
-(4+\A_1\A_2)a_0 +           5a_1 +2b_1=0, && \\
 (5+\A_1\A_2)a_0 -(20+\A_1\A_2)a_1+16a_2
-(12+2\A_1\A_2)b_1                +12b_2=0, &&
\eeqa
\beqa
 \{3m^2+2m+\A_1\A_2\}a_{m-1}
-\{6(m+1)^2-2(m+1)+\A_1\A_2\}a_{m}&&\CR
                 +(3m+5)(m+1)a_{m+1}
         +m(m+\A_1)(m+\A_2)b_{m-1}&&\CR
-(m+1)\{2(m+1)^2-(m+1)+\A_1\A_2\}b_{m}&&\CR
                          +(m+2)(m+1)^2b_{m+1}=0,
\qquad  \hbox{for $m\ge 2$}.
\eeqa
Consequently we obtain the following expressions for 
$(\A_1,\A_2)=({1\/3},{2\/3})$,   
\beqa 
      V_1(u) &=& u + {11\/18}\,u^2 + {109\/243}\,u^3 +{9389\/26244}\,u^4
+\cdots,
\label{sol-z=1-E6}\\
      V_2(u) &=& V_1(u)\log{u} 
                   + {7\/12}\,u^2 + {877\/1458}\,u^3 
+{176015\/314928}\,u^4+\cdots.
\eeqa
For $(\A_1,\A_2)=({1\/4},{3\/4})$ we have
\beqa
 V_1(u) &=& u + {19\/32}\,u^2 + {1321\/3072}\,u^3
+ {22291\/65536}\, u^4+\cdots ,\\
 V_2(u) &=& V_1(u)\log{u} 
              + {39\/64}\,u^2 + {5729\/9216}\,u^3 
+{451495\/786432}\,u^4+\cdots.
\eeqa
For $(\A_1,\A_2)=({1\/6},{5\/6})$ we get
\beqa 
 V_1(u) &=& u +{41\/72}\,u^2 +{6289\/15552}\,u^3 
+ {2122721\/6718464}\, u^4+\cdots,\\
 V_2(u) &=& V_1(u)\log{u} 
            +{31\/48}\,u^2 +{30281\/46656}\,u^3 
+{47918861\/80621568}\,u^4 +\cdots.
\label{sol-z=1-E8}
\eeqa
%

\subsubsection{Torus periods}
\label{torus solution}

Let us now examine torus periods to find the closed form of $V_1(u)$ 
and $V_2(u)$.
The Picard--Fuchs equations for the torus periods are
\beqa
 {\cal L}_{\text{ell}}\varPi_{\text{torus}}=
 \left\{
      \varTheta_z^2-z(\varTheta_z+\A_1)(\varTheta_z+\A_2)
 \right\}    \varPi_{\text{torus}}=0
\label{P-F torus}
\eeqa
whose solutions are given by Meijer G-functions
\beqa 
\varpi_0(z)&=&{\sin{\pi\A_1}\/\pi }
             \,\text{G}_{2,2}^{1,2}
                \left(\left.\ba{cc}\A_1&\A_2\\0&0\ea\right|-z\right), \\
\varpi_1(z)&=&{\sin{\pi\A_1}\/\pi }
             \,\text{G}_{2,2}^{2,2}
                \left(\left.\ba{cc}\A_1&\A_2\\0&0\ea\right|z\right).
\eeqa

Proceeding in parallel with sections \ref{Solutions at $z=0$} and 
\ref{subsec:Solutions at z=infty} we first obtain the solutions
at $z=0$ 
\beqa
\varpi_0(z)&=&{}_2\text{F}_1(\A_1,\A_2 ;1;z), \\
\varpi_1(z)&=&-\varpi_0(z)\log{\left(z\/e^{\beta}\right)}
              -\sum_{n=1}^{\infty}n A(n)\left( B(n)+{1\/n} \right)z^n .
\label{varpi1}
\eeqa

The solutions at $z=\infty$ turn out to be
\beqa
\left(\ba{c}
   \varpi_0^{\infty}(\zeta)\\ 
   \varpi_1^{\infty}(\zeta)
\ea\right)
   =\left(\ba{cc} 
            X_{11}&\ X_{12}\\ X_{21}&\ X_{22}
     \ea\right)
     \left(\ba{c}
           \zeta^{\A_1}\,
           {}_2\text{F}_1\left(\A_1,\A_1;2\A_1;\zeta\right)\\
           \zeta^{\A_2}\,
           {}_2\text{F}_1\left(\A_2,\A_2;2\A_2;\zeta\right) 
     \ea\right) ,\label{Xtorus}
\eeqa
where $\zeta = 1/z$ and
\begin{alignat}{2}
  X_{11}&=&\ {e^{-i\pi \A_1}
              \varGamma(\A_2-\A_1)\/\varGamma(\A_2)^2},
\quad
  X_{12}&=   X_{11}(\A_1\leftrightarrow\A_2),
\CR
  X_{21}&=&\ {\varGamma(\A_1)
              \varGamma(\A_2-\A_1)\/\varGamma(\A_2)},
\quad
  X_{22}&=   X_{21}(\A_1\leftrightarrow\A_2) .
\end{alignat}

As opposite to the case of orbifold/del Pezzo models, 
the  Picard--Fuchs equation around $z=1$ takes the same form 
as the one around $z=0$
\beqa
 \left\{\varTheta_u^2-u(\varTheta_u+\A_1)(\varTheta_u+\A_2)\right\}
 \varPi_{\text{torus}}
  =0,
\eeqa
where $u=1-z$.
Hence its solutions are given by $\varpi_0(u)$ and $\varpi_1(u)$.
Using the Barnes' Lemma \cite[p.289]{WhittakerWatson}, 
\beqa
  &&{}_2\text{F}_1(\A_1,\A_2;1;z) \CR
 &=&  {\sin^2{\pi\A_1}\/\pi^2}
      \int_{-i\infty}^{ i\infty}{ds\/2\pi i}
      \int_{-i\infty}^{ i\infty}{dt\/2\pi i}
               \varGamma(\A_1+t)\varGamma(\A_2+t)
               \varGamma(s-t)\varGamma(-t)\varGamma(-s)(-z)^s \CR
 &=& {\sin^2{\pi\A_1}\/\pi^2}
      \int_{-i\infty}^{ i\infty}{dt\/2\pi i}
               \varGamma(\A_1+t)\varGamma(\A_2+t)
               \varGamma(-t)^2 (1-z)^t,            
       \quad |\arg(-z)|< \pi ,
\label{Barnes' lemma}
\eeqa
we get the connection formulas for torus periods
\beqa 
\varpi_0(z) = {\sin{\pi\A_1}\/\pi}\,\varpi_1(u),\quad
\varpi_1(z) = {\pi\/\sin{\pi\A_1}}\,\varpi_0(u).
\label{a-c of torus}
\eeqa
%

\subsubsection{Solutions based on torus periods}
\label{analytic continuation}

Since ${\cal L}={\cal L}_{\text{ell}}\circ \varTheta_z$,  
the orbifold/del Pezzo periods can be obtained as the logarithmic integral 
of the torus periods. 
In fact, for $|z|<1$ they are related through
\beqa 
\varTheta_z U_1(z)=\varpi_0(z),\quad                  
\varTheta_z U_2(z)=\varpi_1(z).     
\label{overline U}
\eeqa
With the help of this relation and (\ref{a-c of torus}), $U_i(z)$ can be 
analytically continued in the patch $|1-z|<1$.
First we have 
\beqa
 U_1(z) = -{\sin \pi\A_1\/\pi} \int_0^u {du'\/1-u'}\varpi_1(u') + C_1,
\label{U1int-torus}\\
 U_2(z) = -{\pi\/\sin \pi\A_1} \int_0^u {du'\/1-u'}\varpi_0(u') + C_2,
\label{U2int-torus}
\eeqa
where $C_i$ are intgration constants.
Then we assume
\beqa   
U_1(z)&=&A_1 V_1(u)+B_1 V_2(u)+C_1,\CR
U_2(z)&=&A_2 V_1(u)+B_2 V_2(u)+C_2,
\label{analytic continu}
\eeqa
where $V_i(u)$ have been defined in (\ref{omega1}), (\ref{omega2}), 
and $A_i$, $B_i$ are connection coefficients.
Performing the integrals in (\ref{U1int-torus}), (\ref{U2int-torus}) 
we arrive at
\beqa
 V_1(u)&=&
    \sum_{k=0}^{\infty}\sum_{n=0}^{\infty}n A(n){u^{k+n+1}\/k+n+1},
\label{V1 power}\\
 V_2(u)&=&
    V_1(u)\log u\CR
 &+&\sum_{k=0}^{\infty}\sum_{n=0}^{\infty}nA(n)
         \left(  1+B(n)+{1\/n}-{1\/k+n+1}  \right){u^{k+n+1}\/k+n+1}
\label{V2 power}
\eeqa
which indeed agree with (\ref{sol-z=1-E6})--(\ref{sol-z=1-E8}).
We also fix the coefficients $A_i$, $B_i$ as
\begin{alignat}{2}
 A_1 &=&\ -{\sin \pi\A_1\/\pi}(1+\beta),\quad\quad
 B_1 &=     {\sin \pi\A_1\/\pi},             \CR
 A_2 &=&\ -{\pi \/\sin \pi\A_1},\qquad\qquad \quad
 B_2 &= 0. 
\end{alignat}

Our remaining task is to determine the constants $C_i$.
For this we notice that $U_i(1)=C_i$ and $U_i(1)$ themselves can be 
determined by $U_i(1)=U_i^{\infty}(\zeta=1)$.
Eq.(\ref{overline U}) is rewritten as
\beqa
U_2^{\infty}(\zeta)=-\int_0^{\zeta}{d\zeta'\/\zeta'}\varpi_1^{\infty}(\zeta'),
\eeqa
where $U_2^{\infty}(\zeta=0)=0$ has been used.
Substituting here the analytic continuation formula
\beqa
{\varpi_1^{\infty}(\zeta)\/\zeta} = 
{\pi \/ \sin\pi\A_1}\zeta^{-1+\A_1}{}_2\text{F}_1(\A_1,\A_1;1;1-\zeta),
\eeqa
we evaluate \cite{MOY}
\beqa
C_2
&=&-{\pi\/\sin\pi\A_1}\int_0^1 d\zeta\,\zeta^{-1+\A_1}
{}_2\text{F}_1(\A_1,\A_1;1;,1-\zeta)\CR
&=&-\varGamma(\A_2)
    \sum_{n=0}^{\infty}{\varGamma(\A_1+n)^2\/\varGamma(\A_1+n+1)n!}\CR
&=& -{\pi^2\/\sin^2\pi\A_1}.
\eeqa
Therefore we find 
\beqa
 {U_2(1)\/(2\pi i)^2}={1\/\lambda}
\eeqa
which will play a role later.

In a similar vein one can determine $C_1=U_1(1)=U_1^{\infty}(1)$ as follows
\cite{MinahanNem} 
\beqa
 C_1
   &=&  -\int_0^1 {d\zeta\/\zeta} \varpi_0^{\infty}(\zeta) \CR
   &=&  {i\pi \/ \varGamma(\A_1)^2\varGamma(\A_2)^2}
         \int_0^1 {d\zeta\/\zeta} \varpi_1^{\infty}(\zeta)  \CR
   & &  -{\cos \pi \A_1\/ \varGamma(\A_1)\varGamma(\A_2)}
         \int_0^1 {d\zeta\/\zeta} 
             \Bigl( 
                X_{21}\zeta^{\A_1}{}_2\text{F}_1(\A_1,\A_1;2\A_1;\zeta)
               -X_{22}\zeta^{\A_2}{}_2\text{F}_1(\A_2,\A_2;2\A_2;\zeta)
             \Bigr). \CR
\label{integC1}
\eeqa
Here the first term has already been evaluated as above, 
whereas the second term is computed numerically.
The results read
\beqa
 {C_1\/2\pi i}={1\/2}+\left\{\ba{c}
 i\,0.462757882001768178\cdots, \quad \hbox{for ${\vec Z}_3$ (or $E_6$)},\\
 i\,0.610262151883452845\cdots, \quad \hbox{for ${\vec Z}_4$ (or $E_7$)},\\
 i\,0.928067181776930407\cdots, \quad \hbox{for ${\vec Z}_6$ (or $E_8$)}.
\ea\right. 
\label{valueC1}
\eeqa
It is now possible to check that the monodromy matrix at $z=1$ is indeed 
given by (\ref{M_1}).

In view of (\ref{UinfG}), remember that $C_1$ is the value of the 
Meijer G-function at $\zeta =1$
\beqa 
C_1=-{\sin{\pi \A_1}\/ \pi}\,\text{G}_{3,3}^{2,2}
\left({\left.
\ba{ccc}  1 &  1 & 1 \\
        \A_1&\A_2& 0 \ea
\right| -1 }\right).
\eeqa
We wish to point out an amazing relationship of the values of $C_1$ to the
special values of zeta functions in number theory. For this let us introduce
the Hurwitz zeta function 
\beqa
\zeta(s,a)=\sum_{n=0}^\infty {1 \over (n+a)^s}
\eeqa
for $a>0$ \cite{zeta}. It converges absolutely for $\text{Re}\, s>1$ and
reduces to the Riemann zeta function for $a=1$. $\zeta(s,a)$ can be
analytically continued over the complex $s$-plane except for $s=1$ at which 
a simple pole appears. We also introduce the Dirichlet $L$-function
\beqa
L(s,\chi)=\sum_{n=1}^\infty {\chi(n) \over n^s},
\eeqa
where
$\chi(n)$, called the Dirichlet character, obeys $\chi(n+f)=\chi(n)$ with
a positive integer $f$, $\chi(mn)=\chi(m)\chi(n)$ if $m$ and $n$ are prime to
$f$ and $\chi(n)=0$ if $n$ is not prime to $f$. These two zeta functions are
related through
\beqa
L(s,\chi)=f^{-s}\sum_{n=1}^f \chi(n)\, \zeta \bigl(s, {n \over f}\bigr).
\label{Landzeta}
\eeqa

Now, for the ${\vec Z}_3$ (or $E_6$) model, there exists a remarkable relation
proved by Rodriguez Villegas \cite{FRV}
\beqa
\text{Im}\, \bigl({C_1 \over 2\pi i}\bigr)={9\over 2\pi} L'(-1,\chi),
\label{e6zeta}
\eeqa
where $'$ stands for ${d \over ds}$ and $\chi(n)$ has been defined with $f=3$
\beqa
 \chi(n)=\left\{\ba{rl}
  1,& \quad \hbox{$n=1$ mod 3},\\
 -1,& \quad \hbox{$n=2$ mod 3},\\
  0,& \quad \hbox{$n=3$ mod 3}.
\ea\right. 
\label{chi-110}
\eeqa
Namely the $L$-function in (\ref{e6zeta}) reads
\beqa
L(s,\chi)=3^{-s}
\left( \zeta \bigl(s, {1\over 3}\bigr)-\zeta\bigl(s, {2 \over 3}\bigr)\right).
\eeqa
To be convinced, one can check (\ref{e6zeta}) numerically by using the
software package Maple to compute special values of $\zeta '(s,a)$ and 
reproduce (\ref{valueC1}). The proof of (\ref{e6zeta}) is based on the
relation between special values of $L$-function and the Mahler measure
in number theory, which we will discuss further in section 4.4.

For the ${\vec Z}_4$ (or $E_7$) model, we discover by numerical experiment that
\beqa
\text{Im}\, \bigl({C_1 \over 2\pi i}\bigr)={2\over 2\pi} L'(-1,\chi),
\label{e7zeta}
\eeqa
where
\beqa
 \chi(n)=\left\{\ba{rl}
1, & \quad \hbox{$n=1,3$ mod 8},\\
-1,& \quad \hbox{$n=5,7$ mod 8},\\
0, & \quad \hbox{$n=2,4,6,8$ mod 8},
\ea\right. 
\eeqa
and 
\beqa
L(s,\chi)=8^{-s}
\left( \zeta \bigl(s, {1\over 8}\bigr)+\zeta \bigl(s, {3\over 8}\bigr)
-\zeta \bigl(s, {5\over 8}\bigr)-\zeta \bigl(s, {7\over 8}\bigr)\right).
\label{Le7}
\eeqa

For the ${\vec Z}_6$ (or $E_8$) model, we find again by experiment that
\beqa
\text{Im}\, \bigl({C_1 \over 2\pi i}\bigr)={10\over 2\pi} L'(-1,\chi),
\label{e8zeta}
\eeqa
where
\beqa
 \chi(n)=\left\{\ba{rl}
1, & \quad \hbox{$n=1$ mod 4},\\
-1,& \quad \hbox{$n=3$ mod 4},\\
0, & \quad \hbox{$n=2,4$ mod 4},
\ea\right. 
\eeqa
and 
\beqa
L(s,\chi)=4^{-s}
\left( \zeta \bigl(s, {1\over 4}\bigr)-\zeta \bigl(s, {3 \over 4}\bigr)\right).
\label{Le8}
\eeqa
In addition it is seen \cite{Boyd} that
\beqa
\text{Im}\, \bigl({C_1 \over 2\pi i}\bigr)={10\over \pi^2} G,
\eeqa
where $G$ is known as Catalan's constant given by
\beqa
G=\sum_{n=1}^\infty {(-1)^{n-1}\over (2n-1)^2}
 =0.915965594177 \cdots .
\eeqa
Curiously Catalan's constant is ubiquitous in the entropy factors in various
mathematical models \cite{Boyd}.

Although we shall refrain from describing in detail here, the value of $C_1$
for the $E_5$ del Pezzo model is obtained as
\beqa
{C_1 \over 2\pi i}=\half +i\, {4\over 2\pi} L'(-1,\chi),
\label{e5zeta}
\eeqa
where $L(s,\chi)$ is given by (\ref{Le8}) and the expression for 
$\text{Im}\, (C_1/2\pi i)$ is due to \cite{FRV}. 
We see from (\ref{e8zeta}) and (\ref{e5zeta}) that
\beqa
\left[\text{Im}\, \bigl({C_1 \over 2\pi i}\bigr)\right]_{E_8}
\Big/ 
\left[\text{Im}\, \bigl({C_1 \over 2\pi i}\bigr)\right]_{E_5} ={5 \over 2}.
\eeqa
{}From the result of \cite{KlemmZaslow}, on the other hand, this ratio is 
evaluated as $2.50000$ in agreement with ours.

Finally we recall that the value of $\text{Im}\,(C_1/2\pi i)$ is of particular 
interest since it gives the exponent which governs the exponential growth 
of the Gromov--Witten invariants $n(k)$ \cite{candel,MinahanNem,KlemmZaslow}
\beqa
|n(k)| \sim {e^{2\pi \text{Im}\, ({C_1\over 2\pi i}) k} \over k^3 \log^2 k}.
\label{G-W-nk}
\eeqa
It is very intriguing that the special values of zeta functions which are
peculiar to number theory reveal themselves in the property of a significant
set of numbers such as local Gromov--Witten invariants of Fano manifolds.


\section{Mirror maps and modular functions}
\label{sec: Mirror map}
\renewcommand{\theequation}{4.\arabic{equation}}\setcounter{equation}{0}

In this section, first we give the definition of the mirror maps
for the non-compact Calabi--Yau models,
which identifies the periods corresponding 
to the D2-brane and the D4-brane.
The latter receives the quantum corrections   
due to the open string world-sheet instantons,
which is related to the closed string world-sheet instantons  
\cite{OoguriOzYin}. 
Hence the study of the disc instanton effects 
on the D4-brane period in our local Calabi--Yau models
is reduced to that of the Gromov--Witten invariants (of genus zero), 
which have already been done in the literature 
\cite{KlemmMayrVafa,LeMayWar}.

On the other hand, the mirror maps of the $E_{6,7,8}$ 
elliptic curves associated with the local Calabi--Yau models
can be  beautifully described by classical modular functions.
Our second aim in this section is then to elucidate
the relation between  the Gromov--Witten invariants 
of the local Calabi--Yau models and these modular functions. 

Furthermore we find a beautiful link which connects some arithmetic 
properties of local mirror symmetry with a recent topic in number theory;
the Mahler measure and special values of $L$-functions. Describing this
observation is our third aim in this section.

\subsection{Mirror maps for local Calabi--Yau}
\label{subsec: Mirror maps for local Calabi--Yaus}

In this subsection we give the mirror map for orbifolds and del Pezzo models.
In the discussion of mirror symmetry,
it is sometimes convenient to use the {\em unnormalized} modulus parameter
$z_0:=e^{-\beta}z$ instead of $z$.
Let $t_b$, $t$  be the complexified K\"ahler parameters of the orbifold
and the del Pezzo model.
According to \cite{AsGrMo,As} and \cite{LeMayWar}, 
they are given by the solutions of 
the Picard--Fuchs equation of the forms:
\beqa
2\pi i t_b&=& \log(z_0) +\text{O}(z_0),\\
2\pi i t  &=& \log(-z_0)+\text{O}(-z_0),
\label{delPezzo-mirror}
\eeqa
{}from which we can determine the mirror maps as
\beqa
2\pi i t_b &=&  U_1(e^{\beta}z_0)-\pi i =
  \log(z_0) +\sum_{n=1}^{\infty} A(n) \left(e^{\beta}z_0\right)^n,\\
2\pi i t &=&  U_1(e^{\beta}z_0) 
= \log(-z_0) +\sum_{n=1}^{\infty} A(n) \left(e^{\beta}z_0\right)^n, 
\eeqa
that is, $t_b=t-{1\/2}$.
We use the notation $t_b=B_b+iJ_b$ and $t=B+iJ$ to show explicitly 
the physical content of the complexified K\"ahler parameters.
At the orbifold point $z=\infty$, the vanishing of the period
$U_1^{\infty}(\zeta=0)=0$ implies 
\beqa
B_b+iJ_b &=& -\frac12,
\label{J_b}\\
B+iJ     &=& 0,
\eeqa
which means that at the orbifold point,
the orbifold model is described by
a non-singular CFT on the type II string world sheet, 
while the local del Pezzo model by a singular CFT.
Note that the complexified K\"ahler parameter
can also be identified with the central charge of the BPS D2-brane
wrapping around the fundamental two-cycle \cite{GreeneKanter}.

The inversion of the mirror map for the local del Pezzo model
(\ref{delPezzo-mirror}) 
is given by
\begin{align}
E_6: \ z_0&=-e^{2\pi i t}-6\, e^{2\cdot 2\pi it} -9\,e^{3\cdot2\pi i t}
-56\, e^{4\cdot 2\pi i t}+\cdots,\\
E_7: \ z_0&=-e^{2\pi i t}-12\, e^{2\cdot 2\pi i t} -6\,e^{3\cdot 2\pi i t}
-688\, e^{4\cdot 2\pi i t}+\cdots,\\
E_8: \ z_0&=-e^{2\pi i t}-60\, e^{2\cdot 2\pi i t} +1530\,e^{3\cdot 2\pi i t}
-274160\, e^{4\cdot 2\pi i t}+\cdots.
\end{align}

Next we consider the period which represents the D4-brane,
which we denote by $t_d$ and $t_{dP}$ for the orbifold 
and the local del Pezzo model.
In general, all the periods which have $\log^2(z_0)$ with an appropriate
coefficient as the leading term of the large radius limit $z_0\to 0$ 
can be called the D4-brane, 
that is, the definition of the D4-brane period has an ambiguity of
addition of lower dimensional brane charges \cite{GreeneKanter}.  
However, we can uniquely determine $t_b$ and $t_{dP}$ by 
imposing reasonable conditions  on them.    
For the orbifold model, we require that $t_d$ should vanish
at the conifold point $z=1$ \cite{DiacGom}, from  which 
$t_d$ is fixed up to the normalization.
For the local del Pezzo model, on the other hand,
it turns out that $t_{dP}$ should vanish at the orbifold point 
$z=\infty$ \cite{LeMayWar}, 
which leaves the ambiguity of the addition of $t$
to $t_{dP}$. However the form of the central charge at the large radius
region can be used to fix it. 
Finally the normalization factors for the D4-branes $t_d$, $t_{dP}$ 
can be determined by the volume of the twofolds associated with
the local Calabi--Yau models, which we leave to the next section. 
Thus  we arrive at the following results for the {\em unnormalized}
D4-brane periods:
\begin{align}
t_d &=-{U_2(z)\/(2\pi i)^2} + {1\/\lambda},\\
t_{dP}&=-\frac{U_2(z)}{(2\pi i)^2}.
\end{align}
Notice that for the local del Pezzo case, {\em D2- and D4-brane periods are 
given essentially by the Meijer G-functions}.

In the large radius region $|z|<1$ of the orbifold model we obtain 
\beqa
t_d &=& {t_b^2\/2}+{1\/(2\pi i)^2}\left({\pi^2\/2}-\xi\right)+{1\/\lambda}
        +\text{O}(e^{2\pi i t_b})  \CR
    &=& {t_b^2\/2} + {a+b+ab\/24} +\text{O}(e^{2\pi i t_b})
\label{t_dt_b}
\eeqa
corresponding to the exceptional divisor $\text{\bf P}(1,a,b)$ 
in the ${\vec{C}^3/\vec{Z}_m}$ model.
In the $E_{N=6,7,8}$ del Pezzo model, on the other hand,
it follows that 
\beqa
 t_{dP} ={t^2\/2}-{t\/2}+{1\/12}\,{3-N\/9-N}+\text{O}(e^{2\pi i t}),
\label{t_dP-t}
\eeqa
respectively \cite{MOY} .

\subsection{Mirror map for tori}

The mirror map of the torus is 
\begin{equation}
 2\pi i \tau = -{\varpi_1(z)\/\varpi_0(z)}, 
\label{torus-mirror}
\end{equation}
where $\tau$ is the K\"ahler modulus parameter of the torus.
Using the relation (\ref{overline U}) we can show that 
\beqa
\tau = {dt_{dP}\/dt} = t-\frac12+\text{O}(e^{2\pi i t}),
\label{ju-you}
\eeqa
which will play an important role in the investigation of the Gromov--Witten
invariants in the later subsection.

The inversion of the mirror map for $z_0$  
has the following expansion with $q=e^{2\pi i \tau}$:  
\begin{alignat}{2}
E_6:&\  z_0\ & = q&- 15\,q^2+  171\,q^3 -1679\,q^4+ 15054\,q^5+\cdots,
\label{invE6}\\
E_7:&\  z_0\ & = q&- 40\,q^2+ 1324\,q^3 -39872\,q^4+1136334\,q^5+\cdots,
\label{invE7}\\
E_8:&\  z_0\ & = q&-312\,q^2+87084\,q^3 -23067968\,q^4+5930898126\,q^5+\cdots.
\label{invE8}
\end{alignat}
There is an efficient way to obtain the power series expansions above.
First, it is well-known that the inversion of the mirror maps
of $E_{6,7}$ tori (\ref{invE6}), (\ref{invE7}) can be written 
by the {\em Hauptmodul} of the genus zero subgroups 
$\varGamma_0(3)$, $\varGamma_0(2)$ of the modular group
$\varGamma:=\text{PSL}(2;\vec{Z})$,
which are given by the Thompson series $T_{3B}(q)$, $T_{2B}(q)$
\cite{KlemmLercheMayr,LianYau}; see \cite{ConwayNorton} for notations:
\begin{align}
E_6:\  z_0(q)&=\frac{1}{T_{3B}(q)+27}, \qquad 
T_{3B}(q)=\left(\frac{\eta(q)}{\eta(q^3)}\right)^{12},\\
E_7:\  z_0(q)&=\frac{1}{T_{2B}(q)+64}, \qquad
T_{2B}(q)=\left(\frac{\eta(q)}{\eta(q^2)}\right)^{24},
\end{align} 
where $\eta(q)=q^{\frac{1}{24}}\prod_{n\geq 1}(1-q^n)$ is the 
Dedekind eta function.
On the other hand, the inversion for the $E_8$ case (\ref{invE8})
is given by the formal $q$-expansion of the function
\begin{equation}
E_8:\quad z_0(q)=\frac{2}{j(q)+\sqrt{j(q)(j(q)-1728)} },
\label{invE8-j}
\end{equation} 
where $j(q)$ is the $j$-invariant defined by
\beqa
j(q)=\frac{E_4(q)^3}{\eta(q)^{24}}
=\frac{1}{q}+744+196884\,q+21493760\,q^2+864299970\,q^3+\cdots.
\eeqa
Here $E_4(q)$ is the Eisenstein series of weight four, 
also known as the theta function of the $E_8$ lattice
\begin{equation}
E_4(q)
   =  \left(2\frac{\eta(q^2)^2}{\eta(q)}\right)^8
    + \left(\frac{\eta(q)^2}{\eta(q^2)}\right)^8
   =1+240\sum_{n=1}^{\infty}n^3\frac{q^n}{1-q^n}.
\end{equation}
We note that (\ref{invE8-j}) has the following integral representation
\begin{equation}
E_8:\quad z_0(q)=\int_0^q\frac{dq^{'}}{q^{'}}
\frac{E_4(q^{'})^{\frac12}}{j(q^{'})}.
\end{equation} 
Curiously, the following combinations, which can be expressed  
by the Hauptmodul of the genus zero subgroups
$\varGamma_{0}(3)_{+}$, $\varGamma_{0}(2)_{+}$ and $\varGamma$ 
of $\text{PSL}(2;\vec{R})$,
\beqa
E_6: && z_0(1-27z_0)=\frac{1}{T_{3A}(q)+36},\\
E_7: && z_0(1-64z_0)=\frac{1}{T_{2A}(q)+96},\\
E_8: && z_0(1-432z_0)=\frac{1}{j(q)},
\eeqa
coincide with the inversions of the mirror maps of the one-parameter
 family of K3 surfaces: $\text{\bf P}^4[2,3]$, $\text{\bf P}^3[4]$ and 
$\text{\bf P}(1,1,1,3)[6]$ respectively. 
The fundamental period  $\varpi_0$ of the torus
can be written by the modular functions as 
\begin{alignat}{2}
E_6: &  \ & \varpi_0  & = 
\frac{(T_{3B}(q)+27)^{\frac13}}{T_{3B}(q)^{\frac14}}\,
\eta(q)^{2} =
1+6\,q+6\,q^3+6\,q^4+12\,q^7+\cdots,\\
E_7: &  \ & \varpi_0  & = 
\frac{(T_{2B}(q)+64)^{\frac14}}{T_{2B}(q)^{\frac16}}\,
\eta(q)^{2} =
1+12\,q-60\,q^2+768\,q^3-11004\,q^4+\cdots,\\
E_8: &  \  & \varpi_0 & =\!\!\!\!\quad E_4(q)^{\frac14}
 =1+60\,q-4860\,q^2+660480\,q^3-105063420\,q^4+\cdots.
\end{alignat}
%

\subsection{Gromov--Witten invariants}

We begin with the Abel--Liouville theorem \cite{Kohno},
which states that
for the basis $\{\varpi_0,\varpi_1\}$ of the solutions 
of the Picard--Fuchs equation of the $E_{6,7,8}$ tori (\ref{P-F torus}):
\begin{equation}
-\varpi_0(z) \varTheta_z \varpi_1(z)
+\varpi_1(z) \varTheta_z \varpi_0(z)
=\frac{1}{1-z}.
\label{Abel-Liouville}
\end{equation}
Using the mirror map of the torus (\ref{torus-mirror}),
we can recast this equation as \cite[Prop.~4.4]{LianYau2}
\footnote[4]{We note that analogous relations hold in the 
Seiberg--Witten theory
for ${\cal N}\!=\!2$ SU(2) Yang--Mills theory 
with massless fundamental matters
\cite[eq.(2.16)]{KannoYang}.}
%
\begin{equation}
2\pi i\,\varTheta_z \tau=\frac{1}{(1-z)\,\varpi_0(z)^2},
\label{Abel-Liouville2}
\end{equation}
the left hand side of which becomes using 
(\ref{overline U}) and (\ref{ju-you}) 
\begin{equation}
2\pi i z\frac{d\tau}{dz}=2\pi i\, \varTheta_z t\, \frac{d\tau}{dt}
=\varpi_0(z)\, \frac{d\tau}{dt}=\varpi_0(z)\, \frac{d^2 t_{dP}}{dt^2}.
\end{equation}
Therefore  we have the equation for the {\em unnormalized Yukawa coupling}
$Y_{ttt}$ 
\begin{equation}
Y_{ttt}:=\frac{d^2 t_{dP}}{dt^2}=\frac{1}{(1-z)\,\varpi_0(z)^3}
=\frac{1}{(1-z)\, {}_2\text{F}_1(\alpha_1,\alpha_2;1;z)^3}.
\label{yukawa}
\end{equation}
The Yukawa coupling $Y_{ttt}$ may admit two expansions according to 
the two definitions of the mirror maps for the orbifolds and del Pezzos:
\begin{align}
Y_{ttt}&=1-\sum_{k=1}^{\infty} n(k)\, k^3
\,\frac{e^{2\pi i k t}}{1-e^{2\pi ikt}}
\label{Yttt-t}\\
&=1-\sum_{k=1}^{\infty} n_b(k)\, k^3\,
\frac{e^{2\pi i k t_b}}{1-e^{2\pi ikt_b}}.
\label{Yttt-tb}
\end{align}
Since $e^{2\pi i t_b}=-e^{2\pi i t}$, 
the expansion coefficients, which we call the unnormalized
Gromov--Witten invariants, in (\ref{Yttt-t}) and (\ref{Yttt-tb}) 
are related via
\begin{align}
n_b(2k+1)&=-n(2k+1), \nonumber\\
n_b(4k)&=n(4k),\\
n_b(4k+2)&=n(4k+2)+\frac14\,n(2k+1). \nonumber
\end{align}  
This phenomenon was first observed in the relation
between the Gromov--Witten invariants
of the $E_5$ del Pezzo surface and the Hirzebruch surface $\text{\bf F}_0$ 
\cite{LeMayWar};
both models share the Picard--Fuchs operator
${\cal L}_{\text{PF}}=
\{ \varTheta_z^2-z(\varTheta_z+1/2)^2 \} \circ \varTheta_z$, 
but the definitions of the mirror map are different 
just as in our case of the del Pezzo surfaces and orbifolds. 

In terms of the Gromov--Witten invariants,  
the modulus of the torus can be expressed by those of the corresponding
local Calabi--Yau models as
\begin{align}
 q & = - e^{2\pi it}
               \prod_{k=1}^{\infty}(1-e^{2\pi i k t})^{k^2n(k)},\\
   & =  e^{2\pi it_b}
               \prod_{k=1}^{\infty}(1-e^{2\pi i k t_b})^{k^2n_b(k)}.
\label{q-etb}
\end{align}
On the other hand, from (\ref{Abel-Liouville2}) $t_b=t-1/2$ can be obtained 
as the indefinite logarithmic
integration over a combination of the  modular functions
described in the previous subsection:
\begin{equation}
2\pi i t_b=\int \frac{dq'}{q'}\,\left(1-z(q')\right)\, 
\varpi_0\left(z(q')\right)^3.
\label{tb-log-int}
\end{equation}
Explicitly, we have
\begin{alignat}{2}
&E_6: \quad & 
e^{2\pi i t_b}&=q-9\,q^2+54\,q^3-246\,q^4+909\,q^5-2808\,q^6+\cdots,\\
&E_7: \quad & 
e^{2\pi i t_b}&=q-28\,q^2+646\,q^3-13768\,q^4+284369\,q^5-5812884\,q^6
+\cdots,\\
&E_8: \quad & 
e^{2\pi i t_b}&=q-252\,q^2+58374\,q^3-13135368\,q^4+2923010001\,q^5
+\cdots.
\end{alignat}
Comparison of the inversion of these power series and 
(\ref{q-etb}) tells us the invariants $\{n_b(k)\}$ and $\{n(k)\}$. 
The first few values of $n(k)$ may be found, for example,
in \cite{MOY}, and are listed in Table \ref{Ohtake}.

\begin{table}
\begin{center}
\begin{tabular}{|r|r|r|r|} 
\hline
$k$	&$E_6$   		&$E_7$  		&$E_8$\\	
\hline
$1$     &$9$ &$28$ &$252$\\
$2$	&$ -18$  		&$ -136$	&$ -9252$\\
$3$	&$81$ &$1620$ &$848628$\\
$4$	&$ -576$	&$ -29216$	&$ -114265008$\\
$5$	&$5085$	&$651920$ &$18958064400$\\
$6$	&$ -51192$ &$ -16627608$ &$ -3589587111852$\\
$7$	&$565362$	&$465215604$ &$744530011302420$\\
$8$	&$ -6684480$ &$ -13927814272$ &$ -165076694998001856$\\
$9$ &$83246697$	&$439084931544$ &$38512679141944848024$\\
$10$ &$ -1080036450$ &$ -14417814260960$ &$ -9353163584375938364400$\\
$11$ &$ 14483807811$ &$ 489270286160612$ &$ 2346467355966572489025540$\\
$12$ &$ -199613140560$ &$ -17060721785061984$
 &$ -604657435721239536237491472$\\
\hline
\end{tabular}
\end{center}
\caption{Gromov--Witten invariants $n(k)$}
\label{Ohtake}
\end{table}

\subsection{Local mirror from Mahler measure}
Let $P\in \bo{C}[x_1^{\pm},\dots,x_n^{\pm}]$
be a Laurent polynomial in $n$ variables.
The {\em logarithmic Mahler measure} of $P$ 
\cite{Deninger,Boyd,FRV} is defined by
\begin{equation}
m(P)=\frac{1}{(2\pi i)^n}
\int_{\bo{T}}\,\log\left|P(x_1,\dots,x_n)\right|\,
\frac{dx_1}{x_1}\cdots\frac{dx_n}{x_n},
\label{Mahler}
\end{equation}
where 
$\bo{T}=\left\{\,|x_1|=\cdots=|x_n|=1 \right\}$ 
is the standard torus.
If we denote by $\bra P\ket_0$ the constant term in $P$,  
then we have 
\begin{equation}
\left\bra P\right\ket_0=\frac{1}{(2\pi i)^n}
\int_{\bo{T}}\,P(x_1,\dots,x_n)\,
\frac{dx_1}{x_1}\cdots\frac{dx_n}{x_n},
\label{teisu-kou}
\end{equation}
which yields the useful expression for the Mahler measure:
\begin{equation}
m(P)=\operatorname{Re} 
\big\{ \big\bra \log(P)\big\ket_0\big\}.
\end{equation}

Let us consider the Mahler measure of
the one-parameter family of polynomials in two variables 
$P_{\psi}$,
which represents the local mirror geometry of the torus model:
\begin{alignat}{2}
&E_6:\quad &  P_{\psi}(x,y)&=\psi\,xy-(x^3+y^3+1),\nonumber \\ 
&E_7:\quad &  P_{\psi}(x,y)&=\psi\,xy-(x^2+y^4+1),
\label{local-mirror-nosiki} \\
&E_8:\quad &  P_{\psi}(x,y)&=\psi\,xy-(x^2+y^3+1).\nonumber
\end{alignat}
The relation between the modulus parameters reads
$1/z_0=\psi^m$, so that  the sigma model phase corresponds to 
the region 
$|\psi|^m \!>\! e^{\beta}$.
Here we recall that
$m=\{3,4,6\}$ and $e^{\beta}=\{27,64,432\}$ 
for the $E_{\{6,7,8\}}$ family respectively. 

If $|\psi| > 3\ (\geq e^{\beta / m})$,
the following expansion is valid:
\begin{equation}
\log(P_{\psi})-\log(\psi xy)=
\log (1-\psi^{-1}Q)
=-\sum_{n=1}^{\infty}\frac1n \psi^{-n}Q^n,
\label{log-tenkai}
\end{equation}
where 
\begin{align}
E_6:\quad Q(x,y)&=\frac{x^3+y^3+1}{xy},\nonumber \\
E_7:\quad Q(x,y)&=\frac{x^2+y^4+1}{xy},\\
E_8:\quad Q(x,y)&=\frac{x^2+y^3+1}{xy}.\nonumber 
\end{align}
It can be seen that $\bra Q^n \ket_0$ is zero if $n\ne 0 \mod{m}$;
on the other hand 
\begin{align}
E_6:\quad \bra  Q^{3k}\ket_0&=\frac{\varGamma(3k+1)}
{\varGamma(k+1)^3},\nonumber \\
E_7:\quad \bra  Q^{4k}\ket_0&=\frac{\varGamma(4k+1)}
{\varGamma(k+1)^2\varGamma(2k+1)},\nonumber \\
E_8:\quad \bra  Q^{6k}\ket_0&=\frac{\varGamma(6k+1)}
{\varGamma(k+1)\varGamma(2k+1)\varGamma(3k+1)}.\nonumber
\end{align} 
This can be succinctly expressed by $A(k)$ defined in (\ref{B(n)}) as
\begin{equation}
\bra Q^{mk}\ket_0=  e^{k\beta} k A(k).
\label{atai}
\end{equation}

Using (\ref{teisu-kou}),
(\ref{log-tenkai}) and (\ref{atai}), we obtain
the relation between the constant term of $\log(P_{\psi})$ and the
large radius expansion of the period $U_1(z)$ (\ref{U1 z0})
\begin{equation}
\big\bra \log(P_{\psi})\big\ket_0
=\frac{1}{(2\pi i)^2}\int_{\bo{T}}\log(P_{\psi})\,
\frac{dx}{x}\frac{dy}{y}
=-\frac{1}{m}\left(U_1\bigl(\frac{e^{\beta}}{\psi^m}\bigr)-\pi i\right).
\label{fund-local-mirror}
\end{equation}
This is as expected because the middle term is nothing but
the fundamental period of local mirror symmetry 
\cite{ChKlYaZa}.

We see from (\ref{fund-local-mirror})
that in the region  $|\psi|\!>\! 3$,
the Mahler measure of $P_{\psi}$ (\ref{local-mirror-nosiki})
is essentially the same as the real K\"ahler modulus $J$
of the corresponding local Calabi--Yau geometry:
\begin{alignat}{2}
E_6:\quad \frac{3}{2\pi}\, m(P_{\psi}) &=\  &
\operatorname{Im}
\left\{\,\frac{U_1}{2\pi i}\left(\frac{27}{\psi^3}\right)\right\}
&=J\left(\frac{27}{\psi^3}\right),
\label{E6-Mahler} \\
E_7:\quad \frac{4}{2\pi}\,m(P_{\psi}) &=\  &
\operatorname{Im}
\left\{\,\frac{U_1}{2\pi i}\left(\frac{64}{\psi^4}\right)\right\}
&=J\left(\frac{64}{\psi^4}\right),
\label{E7-Mahler}\\
E_8:\quad \frac{6}{2\pi}\,m(P_{\psi}) &=\ &
\operatorname{Im}
\left\{\,\frac{U_1}{2\pi i}\left(\frac{432}{\psi^6}\right)\right\}
&=J\left(\frac{432}{\psi^6}\right).
\label{E8-Mahler}
\end{alignat}

For the $E_6$ model, the K\"ahler modulus 
$J$ in (\ref{E6-Mahler}) can be represented  
as an Eisenstein--Kronecker--Lerch  series \cite{FRV},
which gives the complete expression to (\ref{tb-log-int})
\begin{align}
J\left(\frac{27}{\psi^3}\right)&=
\operatorname{Re}
\left\{\,\operatorname{Im}\tau
+\frac{9}{2\pi}\sum_{n=1}^{\infty}
\sum_{d|n}\chi(d) d^2 \,\frac{q^n}{n}\right\} \nonumber \\
&=\frac{3^{\frac92}\operatorname{Im}\tau}{(2\pi)^3}
\operatorname{Re}
\left\{\underset{n,m\in \bo{Z}}{{\sum}'}
\frac{\chi(n)}{(3m\tau+n)^2(3m{\bar \tau}+n)}
\right\},
\end{align}  
where $\chi$ is the Dirichlet character defined in (\ref{chi-110}).

A quite remarkable relation between the Mahler measures and
the special values of $L$-functions has been found 
\cite{Deninger,Boyd,FRV}.
Needless to say, a  fully  rigorous  treatment of this subject 
is beyond our scope. Nevertheless  we would like to  
quote here a conjecture from \cite[p.~33]{FRV}, which has 
direct relevance to our problem:
For $\psi\in \bo{Z}$,  let $L(s,E_{\psi})$ be 
the Hasse--Weil $L$-function of the corresponding
elliptic curve $E_{\psi}$ defined by (\ref{local-mirror-nosiki}).
Then for all sufficiently large $\psi$, the Mahler measure of 
$P_{\psi}$ coincides with the special value of 
the $L$-function of $E_{\psi}$ up to a multiplication by
a  nonzero rational number:
\begin{equation}
L'(0,E_{\psi})=r_{\psi}\,m(P_{\psi}),\qquad
r_{\psi}\in \bo{Q}^{*}.
\label{conjecture-Villegas}
\end{equation}
It follows immediately that the value of the real K\"ahler 
modulus $J(e^{\beta}/\psi^m)$ of the local Calabi--Yau geometry
with  $\psi$ for which 
the conjecture (\ref{conjecture-Villegas}) is valid
can be given by the special value of the $L$-function of
the elliptic curve $E_{\psi}$.

Take, for example, the $E_8$ model. 
Then the conjecture is rewritten as 
\begin{equation}
J\left(\frac{432}{\psi^6}\right)
=\frac{6}{2\pi}\,\frac{1}{r_{\psi}}\,L'(0,E_{\psi}),
\qquad r_{\psi}\in \bo{Q}^*.
\label{tadano-kakikae}
\end{equation}
In fact, the {\em numerical experiment} for the $E_8$ 
family of the curves by Boyd \cite{Boyd} shows the validity
 of the conjecture (\ref{conjecture-Villegas}) for $3\leq\psi\leq 18$.%
\footnote[5]{Note that $\psi\!=\!2$ is not in  the sigma model phase, 
while the rapid growth of the conductor of the elliptic curve 
$E_{\psi}$ makes it difficult to compute $L'(0,E_{\psi})$ 
for $\psi\!>\!18$.}
Borrowing his data, we list  in Table \ref{J-value}
the values of the real K\"ahler modulus
of our local Calabi--Yau model $J(\frac{432}{\psi^6})$   
as well as the rational numbers $r_{\psi}$
{\em unspecified} in the conjecture.
%
%
\begin{table}
\begin{center}
\begin{tabular}{|r|r|c|} 
\hline
$\psi$	 &$r_{\psi}$   		&$J(\frac{432}{\psi^6})$\\	
\hline
3     &$-4/3$     & 1.03304893002510628669 $\cdots$\\
4     &$-72$      & 1.32141313308322098021 $\cdots$\\
5     & 168       & 1.53628426583345256681 $\cdots$\\
6     &$-216$     & 1.71079907475933497399 $\cdots$\\
7     &$-1152$    & 1.85812606670894012215 $\cdots$\\
8     & 2688      & 1.98568395763630817133 $\cdots$\\
9     & 1440      & 2.09817694280347199839 $\cdots$\\
10    & 10704     & 2.19879724623853723282 $\cdots$\\
11    &$-14400$   & 2.28981592341485331429 $\cdots$\\
12    & 7920      & 2.37290786045027306396 $\cdots$\\
13    & 30888     & 2.44934423568787924171 $\cdots$\\
14    & 7488      & 2.52011284640251294912 $\cdots$\\
15    & 24480     & 2.58599661552298151995 $\cdots$\\
16   &$-155520$   & 2.64762663264546979711 $\cdots$\\
17   &$-139392$   & 2.70551905562125080466 $\cdots$\\
18   & 82368      & 2.76010143509263748236 $\cdots$\\
\hline
\end{tabular}
\end{center}
\caption{Real K\"ahler modulus $J(\frac{432}{\psi^6})$ 
for $E_8$ del Pezzo model.}
\label{J-value}
\end{table}

Now we consider the mirror map of the local Calabi--Yau
model at the discriminant locus  $z\!=\!1$.
The value of the K\"ahler modulus at this point 
$J(1)\!=\!\operatorname{Im}\{C_1/(2\pi i)\}$
is of great importance because it determines the asymptotic 
large $k$ behavior of the Gromov--Witten invariant $n(k)$  
according to (\ref{G-W-nk}).
In this respect we would like to call
$2\pi J(1)\!=\!-\operatorname{Re} C_1$
the {\em entropy} of the local Calabi--Yau model. 
Note that at the discriminant locus the curve 
(\ref{local-mirror-nosiki}) is no longer elliptic by definition.
Correspondingly,  the $L$-function the special value of which yields 
that  of the K\"ahler modulus $J$ at $z\!=\!1$
becomes the Dirichlet one, which we repeat for convenience:
\begin{alignat}{3}
&E_6:\ & J(1)&=\frac{9}{2\pi}L'(-1,\chi_3) &
&=0.462757882001768178 \cdots ,
\label{E6-saikei} \\
&E_7:\ & J(1)&=\frac{2}{2\pi}L'(-1,\chi_8) &
&=0.610262151883452845 \cdots ,
\label{E7-saikei} \\
&E_8:\ & J(1)&=\frac{10}{2\pi}L'(-1,\chi_4) &
&=0.928067181776930407 \cdots ,
\label{E8-saikei}
\end{alignat}   
where 
(\ref{E6-saikei}) is proved in \cite{FRV}  
while (\ref{E7-saikei}) and (\ref{E8-saikei})
are found by our numerical experiment.
It must not be too difficult to prove the latter two equalities
in a rigorous manner.

\subsection{Monodromy matrices}
\label{sec:Monodromy}

Having fixed the mirror maps let us collect here all the monodromy matrices
relevant to our consideration.
For the orbifold models, if we take the basis $(1,t,t_d)$ 
the monodromy matrices, acting on ${}^t(1,t,t_d)$ from the left, 
with integral entries are obtained 
as in Table \ref{table:monodromy of orbifold}.
Using the basis $(1,t_b,t_d)$, which will be adopted when discussing D-brane 
configurations on $\text{\bf P}(1,a,b)$, we have the result in 
Table \ref{table:monodromy of orbifold-integer}.
To be self-contained we also present in Table 
\ref{table:monodromy of the tori} the well-known monodromies for the 
$E_{6,7,8}$ tori acting on ${}^t(\varpi_0,-\varpi_1/(2\pi i))$.
In particular, for $E_6$ and $E_7$, the Picard--Fuchs monodromy generates
$\varGamma_0(3)$ and $\varGamma_0(2)$, respectively.
The monodromy matrices acting on ${}^t(1,t,t_{dP})$ in the dell Pezzo 
models are given in Table \ref{table:monodromy of the del Pezzo}.
We note again that the monodromy matrix $M_{\infty}$ in Tables 
\ref{table:monodromy of orbifold}--\ref{table:monodromy of the del Pezzo} 
obeys $(M_{\infty})^m=I$ for the $\vec{Z}_{m=3,4,6}$ orbifolds 
and the $E_{6,7,8}$ tori as well as del Pezzo surfaces, and
$M_\infty =M_1M_0$.

\begin{table}[p]
\begin{center}
\begin{tabular}{|c||c|c|c|}
\hline
  {}         & $M_0$ & $M_1$ & $M_{\infty}$ \\
\hline
  $\vec{Z}_3$& $\left(\ba{rrr}1&0&0\\ 1& 1& 0\\0&1&1\ea\right)$  &
               $\left(\ba{rrr}1&0&0\\ 0& 1&-3\\0&0&1\ea\right)$  &
               $\left(\ba{rrr}1&0&0\\ 1&-2&-3\\0&1&1\ea\right)$ \\
\hline
  $\vec{Z}_4$& $\left(\ba{rrr}1&0&0\\ 1& 1& 0\\0&1&1\ea\right)$  &
               $\left(\ba{rrr}1&0&0\\ 0& 1&-2\\0&0&1\ea\right)$  &
               $\left(\ba{rrr}1&0&0\\ 1&-1&-2\\0&1&1\ea\right)$ \\
\hline
  $\vec{Z}_6$& $\left(\ba{rrr}1&0&0\\ 1& 1& 0\\0&1&1\ea\right)$  &
               $\left(\ba{rrr}1&0&0\\ 0& 1&-1\\0&0&1\ea\right)$  &
               $\left(\ba{rrr}1&0&0\\ 1& 0&-1\\0&1&1\ea\right)$ \\
\hline
\end{tabular}
\caption{The monodromy in the integral basis $(1,t,t_d)$ 
for the $\vec{Z}_{3,4,6}$ orbifold models.}
\label{table:monodromy of orbifold}
\end{center}
\end{table}

\begin{table}[p]
\begin{center}
\begin{tabular}{|c||c|c|c|}
\hline
  {}         & $M_0$ & $M_1$ & $M_{\infty}$ \\
\hline
  $\vec{Z}_3$& $\left(\ba{rrr}1&0&0\\ 1& 1& 0\\{1\/2}&1&1\ea\right)$  &
               $\left(\ba{rrr}1&0&0\\ 0& 1&-3\\0&0&1\ea\right)$  &
               $\left(\ba{rrr}1&0&0\\-{1\/2}&-2&-3\\{1\/2}&1&1\ea\right)$ \\
\hline
  $\vec{Z}_4$& $\left(\ba{rrr}1&0&0\\ 1& 1& 0\\{1\/2}&1&1\ea\right)$  &
               $\left(\ba{rrr}1&0&0\\ 0& 1&-2\\0&0&1\ea\right)$  &
               $\left(\ba{rrr}1&0&0\\ 0&-1&-2\\{1\/2}&1&1\ea\right)$ \\
\hline
  $\vec{Z}_6$& $\left(\ba{rrr}1&0&0\\ 1& 1& 0\\{1\/2}&1&1\ea\right)$  &
               $\left(\ba{rrr}1&0&0\\ 0& 1&-1\\0&0&1\ea\right)$  &
               $\left(\ba{rrr}1&0&0\\{1\/2}& 0&-1\\{1\/2}&1&1\ea\right)$ \\
\hline
\end{tabular}
\caption{The monodromy in the basis $(1,t_b,t_d)$ 
for the $\vec{Z}_{3,4,6}$ orbifold models.}
\label{table:monodromy of orbifold-integer}
\end{center}
\end{table}

\begin{table}[p]
\begin{center}
\begin{tabular}{|c||c|c|c|}
\hline
  {}         & $M_0$ & $M_1$ & $M_{\infty}$ \\
\hline
  $E_6$&       $\left(\ba{rr} 1& 0\\ 1&1\ea\right)$  &
               $\left(\ba{rr} 1&-3\\ 0&1\ea\right)$  &
               $\left(\ba{rr}-2&-3\\ 1&1\ea\right)$ \\
\hline
  $E_7$&       $\left(\ba{rr} 1& 0\\ 1&1\ea\right)$  &
               $\left(\ba{rr} 1&-2\\ 0&1\ea\right)$  &
               $\left(\ba{rr}-1&-2\\ 1&1\ea\right)$ \\
\hline
  $E_8$&       $\left(\ba{rr} 1& 0\\ 1&1\ea\right)$  &
               $\left(\ba{rr} 1&-1\\ 0&1\ea\right)$  &
               $\left(\ba{rr} 0&-1\\ 1&1\ea\right)$ \\
\hline
\end{tabular}
\caption{The monodromy for the $E_{6,7,8}$ tori.}
\label{table:monodromy of the tori}
\end{center}
\end{table}

\begin{table}[t]
\begin{center}
\begin{tabular}{|c||c|c|c|}
\hline
  {}         & $M_0$ & $M_1$ & $M_{\infty}$ \\
\hline
  $E_6$& $\left(\ba{rrr}1&0&0\\ 1& 1& 0\\0&1&1\ea\right)$  &
               $\left(\ba{rrr}1&0&0\\ -1& 1&-3\\0&0&1\ea\right)$  &
               $\left(\ba{rrr}1&0&0\\ 0&-2&-3\\0&1&1\ea\right)$ \\
\hline
  $E_7$& $\left(\ba{rrr}1&0&0\\ 1& 1& 0\\0&1&1\ea\right)$  &
               $\left(\ba{rrr}1&0&0\\ -1& 1&-2\\0&0&1\ea\right)$  &
               $\left(\ba{rrr}1&0&0\\ 0&-1&-2\\0&1&1\ea\right)$ \\
\hline
  $E_8$& $\left(\ba{rrr}1&0&0\\ 1& 1& 0\\0&1&1\ea\right)$  &
               $\left(\ba{rrr}1&0&0\\ -1& 1&-1\\0&0&1\ea\right)$  &
               $\left(\ba{rrr}1&0&0\\ 0& 0&-1\\0&1&1\ea\right)$ \\
\hline
\end{tabular}
\caption{The monodromy in the basis $(1,t,t_{dP})$ for the $E_{6,7,8}$ 
del Pezzo models.}
\label{table:monodromy of the del Pezzo}
\end{center}
\end{table}


\section{D-branes wrapping a surface}
\label{D-branes wrapping a surface}
\renewcommand{\theequation}{5.\arabic{equation}}\setcounter{equation}{0}

In the previous section we have determined 
how a complexified K\"ahler class of a surface $S$ 
embedded in a non-compact Calabi--Yau threefold $X$ 
depends on a modulus parameter $z$ in the orbifold models 
for which $S=\text{\bf P}(1,a,b)$ with $(a,b)=(1,1),(1,2),(2,3)$, 
and in the local 
del Pezzo models for which $S=E_{6,7,8}$ del Pezzo surfaces.
The result is now employed to discuss D-brane configurations on $S$.
The RR charge vector of D-branes wrapped on $S$ is given by 
\cite{ChYin,MinaMoore}
\beqa
  Q = \text{ch}(V)\,
      \sqrt{\text{Todd}(T_S)\/\text{Todd}(N_S)}  \quad
      \in \bigoplus_{i=0}^2  H^{2i}(S,\vec{Q}), 
\label{QdP}
\eeqa
where $V$ is a vector bundle on $S$ 
(or, more precisely, a coherent ${\cal O}_S$-module), 
$\text{ch}(V)$ is the Chern character; 
$\text{ch}(V)=r(V)+c_1(V)+\text{ch}_2(V)$
and $T_S$ ($N_S$) is the tangent (normal) bundle to $S$.
The BPS central charge then takes the form in the large radius region
\beqa
 Z = -\int_S e^{-J_S}\,\text{ch}(V)\,
      \sqrt{\text{Todd}(T_S)\/\text{Todd}(N_S)}+\cdots,
\label{Z(cd)}
\eeqa
where $J_S$ is a K\"ahler class of $S$ compatible with an embedding 
$S \hookrightarrow X$ 
and the ellipses stand for possible world-sheet instanton corrections.
Notice that in the present embedding, $N_S$ is isomorphic 
to the canonical line bundle $K_S$, and hence $c_1(N_S)=-c_1(S)$.

\subsection{Local del Pezzo models}
\label{subsec:Local del Pezzo models}

The configuration of D-branes on a del Pezzo surface embedded in a Calabi--Yau 
threefold $X$ has been studied in \cite{HI,HIV,MOY}.
Let us begin with presenting some computations 
based on a description of $E_{6,7,8}$ del Pezzo surfaces 
as hypersurfaces in weighted projective space.
Let $S$ denote $E_{6,7,8}$ del Pezzo surfaces. As explained in section 2.4,
$S$ is realized as a hypersurface of degree $(1+a+b)$ 
in $\text{\bf P}(1,1,a,b)$ 
where $(a,b)=(1,1),(1,2),(2,3)$ for $E_{6,7,8}$ respectively.
Let $D$ be a divisor of $\text{\bf P}(1,1,a,b)$ 
isomorphic to $\text{\bf P}(1,a,b)$ 
and denote $\bar D=D\cap S$.
$\bar D$ has the self-intersection
\beqa
\bar D\!\cdot\! \bar D = {{1+a+b}\/ab} = 9-N
\eeqa
for $E_{N=6,7,8}$.
Calculating the total Chern class with the use of the adjunction formula 
one obtains $c_1(S)=\bar D$, and $c_2(S)=(a+b+ab)\bar D\!\cdot\! \bar D$ from 
which the Euler characteristic of $S$, that is,
$\chi(S)=3+N$, can be reproduced.
The calculation of the Todd class yields
\beqa
 \sqrt{\text{Todd}(T_S)\/\text{Todd}(N_S)}  
 = 1 + {1\/2}\bar D + {{15-N}\/12}w_S,
\eeqa
where $w_S= {1\/9-N}\bar D^2$ and $c_1(N_S)=-\bar D$, 
which holds in the present embedding $S \hookrightarrow X$, 
has been utilized.

Since the first Chern class of $S$ is ample, 
we take the K\"ahler class $J_S=t\bar D$ and write down the central charge 
in the large radius limit \cite{MOY}
\beqa
Z&=&-\int_S e^{-t\bar D}\text{ch}(V)
      \sqrt{\text{Todd}(T_S)\/\text{Todd}(N_S)}
    +\text{O}(e^{2\pi it})\CR
 &=&-r(V)\,\bar D\!\cdot\!\bar D\left({t^2\/2}
    -{t\/2}+{1\/12}\,{3-N\/9-N}\right)
    +d(V)\,t -\chi(V) + \text{O}(e^{2\pi i t}),
\label{Z(cdP)}
\eeqa
where $d(V)=c_1(V)\cdot \bar D$ and the Euler characteristic of $V$ is given
by $\chi(V)=r(V)+{1\/2}d(V)+k(V)$ with  $k(V)=\int_S \text{ch}_2(V)$.

At a generic point of the moduli space, the central charge for the local 
del Pezzo models reads
\beqa
 Z= n_4\,\bar D\!\cdot\!\bar D\,t_{dP} + n_2\, t + n_0,
\label{dPcenter}
\eeqa
where $n_i$ are integers.
The model is dual to a theory on a D3-brane probing the affine 
7-brane backgrounds, in view of which $n_i$ are string junction charges 
\cite{MOY}. In the large radius limit it is clear from (\ref{t_dP-t}) 
that (\ref{dPcenter}) reduces to (\ref{Z(cdP)}). Thus, if a BPS state
with the charge vector $(n_0,n_2,n_4)$ survives all the way 
down to the large radius limit at $z=0$ it should admit a description 
in terms of coherent sheaves on $S$ under the relation \cite{HI,MOY}
\beqa
  n_0=-\chi(V),\quad n_2=d(V),\quad n_4=-r(V).
\label{dPn=c}
\eeqa
It also follows that $\bar D\!\cdot\! \bar D\, t_{dP}$ gives a  
normalized central charge of a D4-brane. 
A bundle (or sheaf in general)
$V$ corresponds to a D-brane with the positive orientation
if  $r(V)>0$, or $r(V)=0$ and $d(V)>0$, or
$r(V)=d(V)=0$ and $\chi(V)<0$, and otherwise to 
a D-brane with the opposite orientation, which we call
a $\overline{\text{D}}$-brane

The homology $H_2(S,\vec{Z})$ of an $E_N$ del Pezzo surface is spanned by a 
generic line $\ell$ in $\text{\bf P}^2$ and the exceptional divisors 
$e_1,\dots ,e_N$ of the
blown-up points. The degree zero sublattice of $H_2(S)$ is isomorphic to the
$E_N$ root lattice with the simple roots; 
$\vec{\alpha}_i=e_i-e_{i+1}\, (1\leq i \leq N-1)$ 
and $\vec{\alpha}_N=\ell -e_1-e_2-e_3$. Then the first Chern
class $c_1(V)$ has the orthogonal decomposition \cite{MOY}
\beqa
c_1(V)={d(V) \over 9-N}\,\bar D +\sum_{i=1}^N\lambda_i(V)\, \vec{w}^i,
\label{dP1st}
\eeqa
where $\vec{w}^i\!\cdot\! \vec{\alpha}_j=-\delta^i_j$ and 
$\bar D\!\cdot\! \vec{w}^i=0$. 
Thus the D2-brane charge is specified not only by the degree $d(V)$ but also 
by the Dynkin label $\{ \lambda_i\}$ of a representation of $E_N$. 
If we turn on all the K\"ahler parameters associated 
with the exceptional divisors, the
central charge formula (\ref{dPcenter}) will be modified so as to contain 
the full dependence on $\{ \lambda_i\}$. 
The second Chern class $c_2(V)$ is now 
evaluated from (\ref{dPn=c}) and (\ref{dP1st}) to be
\beqa
\int_S c_2(V)=n_0+{n_2 \over 2}+
\half \left({n_2^2\over 9-N}-\lambda\!\cdot\!\lambda \right)-n_4.
\label{dP2nd}
\eeqa
Eqs. (\ref{dPn=c}) and (\ref{dP2nd}) enable us to translate the charge
vector $(n_0,n_2,n_4)$ into the sheaf data (modulo the $E_N$ representation).

At $z=\infty$, the $E_{N=6,7,8}$ del Pezzo model exhibits a  
$\vec{Z}_{3,4,6}$ symmetry, respectively.
Since $t=t_{dP}=0$ at $z=\infty$, a BPS state with $n_0=0$ becomes massless,
but a state with $n_0\not= 0$ massive. Let us present typical examples of
$\vec{Z}_{3,4,6}$ orbits of BPS states. In view of a D3-probe theory
\cite{MOY}, we observe that a state with $(n_0,n_2,n_4)=(1,0,1)$ is
BPS, $E_N$ singlet and exists everywhere in the moduli space. 
In fact, according to 
(\ref{dPn=c}), this state is identified with a $\overline{\text{D}4}$-brane
corresponding to $-{\cal O}$ with ${\cal O}$ being the trivial line bundle.
At $z=\infty$, the state $(1,0,1)$ remains massive and its $\vec{Z}_m$ orbits
are constructed by the $\vec{Z}_{3,4,6}$ action on the charge vector 
\beqa
(n_0,n_2,n_4)
\rightarrow
(n_0,n_2,n_4)
\left(\ba{ccc}
             1 & 0   &  0\\
             0 & N-8 & -1\\
             0 & 9-N &  1 \ea \right).
\label{Monodromy-del-Pezzo}
\eeqa
This has been obtained from Table \ref{table:monodromy of the del Pezzo}
by noting that the monodromy matrices 
acting on the periods ${}^t(1,t,t_{dP})$ by left multiplication act on the 
charge vector $(n_0,n_2,n_4)$ by right multiplication. 
We then have the $E_N$ singlet massive $\vec{Z}_m$ orbits 
associated with the state $(1,0,1)$ and corresponding D-brane
configurations as follows:
\begin{itemize}
\item
$E_6$ del Pezzo
\beqa
(1,0,1) &\rightarrow& \overline{\text{D}4} ,          \CR
(1,3,1) &\rightarrow& \overline{\text{D}4} + \text{D}2,\CR
(1,-3,-2)&\rightarrow&         2{\text{D}4} + \overline{\text{D}2}
                                                + 3\text{D}0.
\eeqa
\item
$E_7$ del Pezzo
\beqa
(1,0,1) &\rightarrow& \overline{\text{D}4} ,                      \CR
(1,2,1) &\rightarrow& \overline{\text{D}4} + \text{D}2            ,\CR
(1,0,-1) &\rightarrow&           \text{D}4  +2\text{D}0             \CR
(1,-2,-1)&\rightarrow&           \text{D}4  + \overline{\text{D}2}
                                                + 2\text{D}0.
\eeqa
\item
$E_8$ del Pezzo
\beqa
(1,0,1) &\rightarrow& \overline{\text{D}4} ,                      \CR
(1,1,1) &\rightarrow& \overline{\text{D}4} + \text{D}2            ,\CR
(1,1,0) &\rightarrow&           \text{D}2  +2\text{D}0            ,\CR
(1,0,-1) &\rightarrow&           \text{D}4  +2\text{D}0            ,\CR
(1,-1,-1)&\rightarrow&           \text{D}4  +\overline{\text{D}2}
                                           +2\text{D}0            ,\CR
(1,-1,0) &\rightarrow& \overline{\text{D}2}             .
\eeqa
\end{itemize}
Note that every D2 (or $\overline{\text{D}2}$)-brane in the above is
 $E_N$ singlet.
It will be very interesting to have a proper interpretation of these 
configurations in terms of vector bundles on del Pezzo surfaces.

Finally let us remark how the monodromy action on the periods induces
the corresponding action on a vector bundle. As just mentioned above,
we know how the monodromy acts on the charge vector, and hence 
we can convert the large radius monodromy action
on the periods to that on the vector bundle under the identification
(\ref{dPn=c}). The result is
\beqa
\text{ch}(V) \rightarrow \text{ch}(V)\,e^{-\bar{D}},
\eeqa
which is in accordance with the fact that the large radius monodromy 
$t\rightarrow t+1$ is induced by a shift of the B-field; $B\rightarrow B+1$.
Similarly the monodromy at $z=1$ leads to 
\beqa
\text{ch}(V) \rightarrow \text{ch}(V)+\int_S\text{ch}(V)\bar{D}.
\eeqa
This is understood to be performed along a loop which is based at the point
$z=0$ (the large radius limit) and encircles the discriminant locus at
$z=1$ \cite{Horja,hosono}. See \cite{DiaconescuRomelsb} for a related
observation in the case of an elliptically fibered Calabi--Yau model.

\subsection{Orbifold models}
\label{subsec:Orbifold models}

Let us next turn to the orbifold models. As we have described in section 2.1,
the blown-up orbifold $\blowup{3,4,6}$ has an exceptional divisor 
$\text{\bf P}(1,a,b)$ with $(a,b)=$ $(1,1)$, $(1,2)$, $(2,3)$, respectively.
In this section
we consider D-branes wrapped on $S=\text{\bf P}(1,a,b)$.
When applying (\ref{QdP}) to the computation of D-brane charges 
one should take into account that the background B-field is turned on 
in the orbifold model as shown in (\ref{J_b}). 
Following \cite{DiacGom, DouglasFiolRomelsb} we assume that 
the $B$-dependence of $\text{ch}(V)$ will cancel out the factor $e^{c_1(S)/2}$ 
appearing in the relation 
\beqa
  \sqrt{\text{Todd}(T_S)\/\text{Todd}(K_S)} 
  = e^{\half c_1(S)}\sqrt{\widehat A(T_S)\/\widehat A(K_S)}
\label{Q}
\eeqa
so that the RR charge vector is read off from
\beqa
  Q =\text{ch}(V)\,
      \sqrt{\widehat A(T_S)\/\widehat A(K_S)}.
\eeqa

Let us set the K\"ahler class 
$J_S=t_b D$, where $D$ is the ample generator of divisors of $S$,
then the classical central charge (\ref{Z(cd)}) takes the form
\beqa
Z=-\int_S e^{-t_bD}\,\text{ch}(V)\,
     \sqrt{\widehat A(T_S)\/\widehat A(K_S)}+\text{O}(e^{2\pi it_b}).
\label{clacenter}
\eeqa
The quantum central charge, on the other hand, is expressed in terms of 
the periods as
\beqa
Z(n_0,n_2,n_4) = n_4\,D\!\cdot\! D\,t_d + n_2\,t_b + n_0,
\label{Z(n)}
\eeqa
where $n_i$ are not necessarily integral. We now wish to show that, 
in the large radius limit, (\ref{clacenter}) is precisely recovered from 
(\ref{Z(n)}). For this we first give the self-intersection of $D$
\beqa
 D\!\cdot\! D = {1\/ab} = 1,\; {1\/2},\; {1\/6}
\eeqa
for $\vec{C}^3/\vec{Z}_{3,4,6}$. Next, using the naive adjunction formula 
we obtain 
\beqa
 \sqrt{\widehat A(T_S)\/\widehat A(K_S)} 
   =  1 + {1\/24}\,\tilde{\chi}(S)\,w_S, 
\eeqa
where $\tilde{\chi}(S)=(a+b+ab)/(ab)$ and $w_S=abD^2$.
The classical central charge (\ref{clacenter}) thereby turns out to be
\beqa
Z  &=&-{1\/2}r(V)\,D\!\cdot\! D\,t_b^2  +d(V)\,t_b
       -{1\/24}r(V)\tilde{\chi}(S)-k(V) +\text{O}(e^{2\pi it_b})   \CR
&=& -r(V)\, D\!\cdot\! D \left( {t_b^2 \over 2}+{a+b+ab \over 24}\right)
    +d(V)\, t_b-k(V) +\text{O}(e^{2\pi it_b}),
\label{Z(c)}
\eeqa
where $d(V)=c_1(V)\!\cdot\! D$ and $k(V)=\int_S\text{ch}_2(V)$. 
It is clearly seen that if we put
\beqa
 n_0=-k(V),\quad n_2=d(V),\quad n_4=-r(V),
\label{n=c}
\eeqa
(\ref{Z(c)}) coincides with (\ref{Z(n)}) by virtue of (\ref{t_dt_b})
in the large radius region.

In (\ref{Z(n)}), thus, $D\!\cdot\! D\,t_d$ plays a role of 
the normalized central 
charge of a D4-brane. Since we have $c_1(V)=ab\, d(V)D$, the second Chern class
is obtained as
\beqa
  \int_S c_2(V) = n_0+{1\/2}ab\,n_2^2.
\label{c2=n}
\eeqa
Using (\ref{n=c}) and (\ref{c2=n}) one can convert the orbifold charges 
$(n_0,n_2,n_4)$ into the sheaf data in the large radius region. When doing
this, the data with negative $r(V)$ as well as $r(V)=0$ is treated as in the
case of local del Pezzo models.

Let us concentrate on the orbifold point $z=\infty$.
Since $t_d(z=\infty)={1\/\lambda} (= {1\/3},{1\/2},1)$ we have a particular 
value of the central charge
\beqa
Z(0,0,1) = D\!\cdot\! D\,{1\/\lambda} ={1 \over ab\lambda}={1\/m}
\eeqa
for the $\vec{C}^3/\vec{Z}_{m=3,4,6}$ models.
This is regarded as $1/m$ of the mass of a D0-brane.
Therefore the configuration $(0,0,1)$ is identified with a fractional brane.
At the orbifold point there exists a $\vec{Z}_m$ quantum symmetry.
Following the del Pezzo case one can read off from 
Table \ref{table:monodromy of orbifold-integer}
the $\vec{Z}_m$ action on the charge vector
\beqa
(n_0,n_2,n_4)
\rightarrow
(n_0,n_2,n_4)
\left(\ba{ccc}
             1 & 0 & 0 \\
             1-{\lambda\/2} &    1-\lambda & -m\\
             {\lambda\/2m} &  {\lambda\/m} & 1 \ea \right).
\label{Monodromy-orbif}
\eeqa
Thus the $\vec{Z}_3$ orbit of fractional branes in the 
$\vec{Z}_3$ orbifold model reads
\beqa
 (0,0,1) \rightarrow \bigl({1\/2},1,1\big) \rightarrow \bigl({1\/2},-1,-2\big).
\eeqa
For the $\vec{Z}_4$ orbifold we have the $\vec{Z}_4$ orbit
\beqa
 (0,0,1) \rightarrow \bigl({1\/4},{1\/2}, 1\big) 
         \rightarrow \bigl({1\/2},     0,-1\big) 
\rightarrow \big({1\/4},-{1\/2},-1\big).
\eeqa
Likewise the $\vec{Z}_6$ orbifold model has the $\vec{Z}_6$ orbit of 
fractional branes
\beqa
 (0,0,1) \rightarrow \bigl({1\/12}, {1\/6}, 1\big) 
         \rightarrow \bigl( {1\/4}, {1\/6}, 0\big) 
\rightarrow \bigl({1\/3},0,-1\big)
         \rightarrow \bigl( {1\/4},-{1\/6},-1\big) 
\rightarrow \bigl({1\/12},-{1\/6},0\big).
\eeqa

These fractional branes are constructed as the boundary states of the 
$\vec{C}^3/\vec{Z}_m$ orbifold CFT at $z=\infty$ \cite{DiacGom}.
If we assume that these BPS states are stable in the large radius limit, 
they should be described as coherent sheaves on $S$.
The states in the $\vec{Z}_m$ orbit are then identified with 
the corresponding D-brane configurations by the use of (\ref{n=c}),
(\ref{c2=n}). Corresponding to the $\vec{Z}_m$ orbits, we get 
the following D-brane configurations:
\begin{itemize}
\item
$\vec{Z}_3$ orbifold 
\beqa
 (0, 0, 1)&\rightarrow &\ \overline{\text{D}4},                    \CR
 \bigl({1\/2}, 1, 1\big)
           &\rightarrow &\ \overline{\text{D}4}+\text{D}2,           \CR
 \bigl({1\/2},-1,-2\big)
           &\rightarrow &\         2 \text{D}4 + \overline{\text{D}2}
+\text{D}0 .
\label{Z3-orb}
\eeqa
\end{itemize}
Here the first two configurations are identified with $-{\cal O}$, 
$-{\cal O}(-1)$, where ${\cal O}$, ${\cal O}(-1)$ are the trivial 
and the tautological line bundles on $\text{\bf P}^2$, 
whereas the third one is 
a rank two exceptional bundle on $\text{\bf P}^2$ \cite{DiacGom}. 
We will review
exceptional bundles on $\text{\bf P}^2$ in Appendix A.
\begin{itemize}
\item
$\vec{Z}_4$ orbifold
\beqa
 ( 0, 0, 1) &\rightarrow &\overline{\text{D}4},                          \CR
 \bigl({1\/4}, {1\/2}, 1\big)
             &\rightarrow &\overline{\text{D}4} + \text{D}2,             \CR
 \bigl({1\/2},   0,-1\big) 
             &\rightarrow &\text{D}4                 +{1\/2}\text{D}0,\CR
 \bigl({1\/4},-{1\/2},-1\big) 
             &\rightarrow &\text{D}4  + \overline{\text{D}2} +{1\/2}\text{D}0.
\eeqa
\item
$\vec{Z}_6$ orbifold
\beqa
 ( 0, 0, 1)&\rightarrow &\overline{\text{D}4},                         \CR
 \bigl({1\/12}, {1\/6}, 1\big)
            &\rightarrow &\overline{\text{D}4}+\text{D}2,                \CR
 \bigl({ 1\/4}, {1\/6}, 0\big)
            &\rightarrow &                    \text{D}2 +{1\/3}\text{D}0,\CR
 \bigl({ 1\/3},    0,  -1\big)
            &\rightarrow &          \text{D}4           +{1\/3}\text{D}0,\CR
 \bigl({1\/4},-{1\/6},-1\big)
            &\rightarrow &          \text{D}4 + \overline{\text{D}2}
                                                       +{1\/3}\text{D}0,\CR
 \bigl({1\/12},-{1\/6}, 0\big)
            &\rightarrow &                   \overline{\text{D} 2}.
\eeqa
\end{itemize}
In the $\vec{Z}_4$ and $\vec{Z}_6$ cases, D-branes wrap $\text{\bf P}(1,1,2)$
and $\text{\bf P}(1,2,3)$ respectively.
Remember that $\text{\bf P}(1,1,2)\simeq\text{\bf P}^2/\vec{Z}_2$ and 
$\text{\bf P}(1,2,3)\simeq \text{\bf P}^2/\vec{Z}_2\times\vec{Z}_3$. 
Namely D-branes 
are on orbifolds with quotient singularities yet to be resolved.
This may result in the fractional values of the second Chern class 
we have observed in the above $\vec{Z}_4$ and $\vec{Z}_6$ orbits.

Finally we note that the large radius monodromy acts on the Chern 
character as
\beqa
 \text{ch}(V) \rightarrow \text{ch}(V)\,e^{-D},
\label{ch-ch*e(-D)}
\eeqa
while under the monodromy at $z=1$ one has
\beqa
 \text{ch}(V)\rightarrow \text{ch}(V)+ m\int_S\text{ch}(V) D,
\eeqa
where $mD=c_1(S)$ is the first Chern class of 
$\text{\bf P}(1,a,b)$ with $m=1+a+b$.

\subsection{Monodromy invariant  intersection form}
%
Let  $\iota\colon S\hookrightarrow X$ 
be an embedding of a surface $S$ in a Calabi--Yau threefold $X$.
We then have the direct image map $\iota_*$ from
the coherent ${\cal O}_S$-modules to
the coherent ${\cal O}_X$-modules.
The canonical intersection form on the vector bundles on $X$ 
is given  by
\begin{align}
I_X(W_1,W_2)&=\int_X\,
\text{ch}(W^{*}_1)\,\text{ch}(W_2)\,\text{Todd}(T_X),
\nonumber \\
&=-I_X(W_2,W_1),
\end{align}
which can be extended  
to an anti-symmetric intersection form on the coherent 
${\cal O}_X$-modules using locally-free resolutions of them.

The intersection form on the vector bundles on $S$ 
induced from that on the ambient Calabi--Yau threefold 
$X$ by the embedding $\iota\colon S\hookrightarrow X$ reads \cite{MOY}
\begin{equation}
A_S(V_1,V_2):=I_X(\iota_*V_1,\iota_*V_2)=
r(V_1)\,d(V_2)-r(V_2)\,d(V_1),
\end{equation}
where $d(V)=c_1(V) \cdot c_1(S)$ is the degree of the bundle.
Note that for $S=\text{\bf P}(1,a,b)$,
$d(V)$ here is $m$ times larger than  
that in the preceding subsection.

$A_S$ does not depend on the detail of the embedding data,
but only on the intrinsic geometry of $S$.
More importantly, 
it is easily verified that 
$A_S$ defines a {\em monodromy invariant} intersection form 
on the D-branes   
both on the $E_{6,7,8}$ del Pezzo surfaces
and on the exceptional divisors $\text{\bf P}(1,a,b)$
of the $\bo{Z}_{3,4,6}$ orbifolds.  

\vskip6mm\noindent
{\bf Acknowledgements}

\vskip2mm
S.K.Y. would like to thank the participants of Summer Institute 2000 at
Yamanashi, Japan, for their interest in the present work, especially T. Eguchi,
K. Hori, H. Kanno, A. Kato and T. Kawai for stimulating discussions.
We would like to thank Y. Ohtake for useful discussions.
The research of K.M. and S.K.Y. was supported in part by Grant-in-Aid for 
Scientific Research 
on Priority Area 707 ``Supersymmetry and Unified Theory of Elementary 
Particles'', Japan Ministry of Education, Science and Culture.

\newpage
\appendix
\renewcommand{\thesection}{}
\section{\!\!\!\!\!\!\!Appendix~A \  
Exceptional bundles on $\protect\text{\bf P}^{\protect\text{\bf 2}}$}
\renewcommand{\theequation}{A.\arabic{equation}}\setcounter{equation}{0}
The intersection pairing $\chi_{\Ptwo}$
on vector bundles on $\Ptwo$ is defined by
\begin{align}
\chi_{\Ptwo}(V_1,V_2)&=\sum_{i=0}^2\, (-1)^i 
\dim H^i(\Ptwo, \Hom(V_1,V_2)),           \nonumber \\
&=r_1\,r_2+r_1\,k_2+r_2\,k_1-d_1\,d_2+\frac32 (r_1\,d_2-r_2\,d_1),
\label{intersection}
\end{align}
where $\Hom(V_1,V_2)\cong V_1^*\otimes V_2$
is the homomorphism bundle,
%
%
and we have used the Riemann--Roch formula \cite{MOY}
with the abbreviated notation:
$r_{1,2}=r(V_{1,2})$,
$d_{1,2}=d(V_{1,2})$,
$k_{1,2}=k(V_{1,2})$
understood.
In particular, the self-intersection of the bundle $V$ becomes
\begin{equation}
\chi_{\Ptwo}(V,V)=\sum_{i=0}^2\, (-1)^i \dim H^i(\Ptwo, \End(V))
=r^2+2\,rk-d^2.
\label{self-intersection}
\end{equation}
It must not be confused with the Euler characteristic of $V$ defined by
\begin{equation}
\chi(V)=\sum_{i=0}^2\, (-1)^i \dim H^i(\Ptwo, V)
=r+\frac32\,d+k.
\end{equation}

We also introduce two other invariants of $V$, that is,
the slope $\mu(V)$ 
and the (normalized) discriminant $\Delta(V)$,
which must be positive for $V$ to be stable: 
\begin{align}
\mu(V)&=\frac{d}{r}, \\
\Delta(V)& 
=\frac12\left(\frac{d}{r}\right)^2-\frac{k}{r},
\end{align}
as well as the polynomial 
$P(z)=1/2(z+1)(z+2)$ for convenience.
We can then easily verify the following
\begin{align}
\chi(V)&=r\big(P(\mu)-\Delta\big),\nonumber \\
\chi(V_1,V_2)&=r_1r_2 \big(P(\mu_2-\mu_1)-\Delta_1-\Delta_2\big).
\end{align}

A vector bundle $E$ on $\Ptwo$ is called {\em exceptional} if
$$
H^0(\Ptwo,\End(E))\cong \bo{C},\quad
H^1(\Ptwo,\End(E))=0, \quad
H^2(\Ptwo,\End(E))=0.
$$ 
It is known that each exceptional bundle is stable, that is,
its slope is greater than that of any coherent subsheaf of it,
and has no moduli, which means that the complex structure
of an exceptional bundle is uniquely determined by 
its topological invariant $(r,d, k)$.
In fact  if $E$ is exceptional, then 
$k$ is not an independent degree of freedom but is written as 
$k=(1+d^2-r^2)/(2r)$ 
because 
\begin{equation}
\chi_{\Ptwo}(E,E)=1
\end{equation}
by definition.   
The remaining two $(r(E),d(E))$ 
must be {\em mutually prime} according to
the formula (\ref{self-intersection}).
Therefore we have seen that
{\em an exceptional bundle} $E$ {\em is uniquely
determined by its slope} $\mu(E)=d/r$.
It is also easy  to see that if $E$ is exceptional then
its discriminant reads
\begin{equation}
0<\Delta(E)=\frac12\left(1-\frac{1}{r^2}\right) < \frac12.
\end{equation}

The exceptional bundles on $\Ptwo$ are completely classified
in \cite{DrezetLePotier}.
Because 
both the Peccei--Quinn symmetry $B\to B+1$ 
discussed in the preceding subsection, 
which operates as  $E\to E(-1)$ so that
$\mu\to \mu-1$,
and the duality transformation
$E\to E^{*}$, which results in 
$\mu\to -\mu$,
preserve the endmorphism bundle $\End(E)$,
it suffices to list the rational numbers 
corresponding to the slopes of the exceptional bundles
in the fundamental domain $[0,1/2]$.

In order to state the result in \cite{DrezetLePotier},
we must first introduce some notations closely following them.
For $\A\!\in\!\bo{Q}$, 
the rank of it, which we denote by  $r_{\A}$,
is the least positive number such that $\A\,r_{\A}\in\!\bo{Z}$.
We also define its discriminant and Euler number by
\begin{equation*}
\Delta_{\A}=\frac12\left(1-\frac{1}{r_{\A}^2} \right),\qquad
\chi_{\A}=r_{\A} \left( P(\A)-\Delta_{\A}\right).
\end{equation*}
For $\A,\B\in\!\bo{Q}$, such that $\B-\A-3\ne 0$, we define a third 
element of $\bo{Q}$ by
\begin{equation*}
\A \circ \B :=\frac12(\A+\B)
+\frac{\Delta_{\B}-\Delta_{\A}}{3+\A-\B}.
\end{equation*}

Let ${\cal D}$ be the subset of $\bo{Q}$ defined by
$$
{\cal D}=\left.\left\{\,\frac{n}{2^q}\,\right|\,
n\in \bo{Z},q\in \bo{N}\cup \{0\} \right\}.
$$
We can define the map $\varepsilon\colon {\cal D}\to \bo{Q}$
uniquely by the requirements:
$\varepsilon(n)=n$ for $n\in \bo{Z}$, and  
$$
\varepsilon\left(\frac{2m+1}{2^{q+1}}\right)=
\varepsilon\left(\frac{m}{2^q}\right) \circ
\varepsilon\left(\frac{m+1}{2^q}\right).
$$
It follows immediately that 
$\varepsilon$ is strictly increasing function,
$\varepsilon(\A+n)=\varepsilon(\A)+n$ for $n\in \bo{Z}$,
$\varepsilon(-\A)=-\varepsilon(\A)$, and 
if $\A\in {\cal D}$, then  $r_{\varepsilon(\A)}\geq r_{\A}$.

The fundamental result of \cite{DrezetLePotier} is that
the set of exceptional bundles on $\Ptwo$ is identified
by their slopes with the subset 
$\operatorname{Im}(\varepsilon)=
\operatorname{Im}\left(\varepsilon\colon {\cal D}\to \bo{Q}\right)$
of $\bo{Q}$. 
Note that from the property of the map $\varepsilon$,
the slope $\mu$ of each exceptional bundle on $\Ptwo$  
with $r<2^{q+1}$ can be put in the finite set
\begin{equation}
\left.\left\{\, \varepsilon\left(\frac{m}{2^q}\right) \,\right|\, 
1\leq m \leq 2^{q-1} \right\}
\subset \bo{Q}\cap \left[0,\frac12\right],
\end{equation}
if we use the symmetries $\mu\to \mu-1$ 
and $\mu\to -\mu$ discussed above. 

 Searching for the elements of 
$\operatorname{Im}(\varepsilon)\cap [0,1/2]$ 
with $2\!\leq\!r\!<\!64$, for example, 
we find, in addition to  $(r,d,k)=(2,1,-1/2)$,
which is the dual of the rank two bundle appeared in the
$\bo{Z}_3$-orbit of the fractional branes (\ref{Z3-orb}),
the four higher rank exceptional bundles \cite{Rudakov}:
\begin{equation}
(r,d,k)= 
(5,2,-2),\ 
\big(13,5,-\frac{11}{2}\big),\  
(29,12,-12),\         
\big(34,13,-\frac{29}{2}\big).
\end{equation}

\newpage


\begin{thebibliography}{99}

\bibitem{Douglas}
See, for a recent review,
M.R.~Douglas,
Topics in D-geometry,
Class.~Quant.~Grav. {\bf 17} (2000) 1057--1070,
hep-th/9910170.

\bibitem{BDLR}
I.~Brunner, M.R.~Douglas, A.~Lawrence and C.~R\"omelsberger,
D-Branes on the Quintic,
JHEP {\bf 0008} (2000) 015,
hep-th/9906200.

\bibitem{DiacGom} D.-E.~Diaconescu and J.~Gomis,
Fractional Branes and Boundary States in Orbifold Theories,
hep-th/9906242.

\bibitem{DiaconescuRomelsb}D.-E.~Diaconescu and C.~R\"omelsberger,
D-Branes and Bundles on Elliptic Fibrations,
Nucl.~Phys.~{\bf B574} (2000) 245--262,
hep-th/9910172.

\bibitem{KasteLercheLutkenWalcher}
P.~Kaste, W.~Lerche, C.A.~L\"utken and J.~Walcher,
D-Branes on K3-Fibrations,
Nucl.~Phys.~{\bf B582} (2000) 203--215,
hep-th/9912147.

\bibitem{Scheidegger} E.~Scheidegger,
D-Branes on Some One- and Two-Parameter Calabi--Yau Hypersurfaces,
JHEP {\bf 0004} (2000) 003,
hep-th/9912188.

\bibitem{GrLa} B.R.~Greene and C.I.~Lazaroiu,
Collapsing D-Branes in Calabi--Yau Moduli Space: I,
hep-th/0001025

\bibitem{Lazaroiu}
C.I.~Lazaroiu,
Collapsing D-Branes in One-Parameter Models and Small/Large Radius Duality,
hep-th/0002004.

\bibitem{DouglasFiolRomelsberger}
M.R.~Douglas, B.~Fiol and C.~R\"omelsberger,
Stability and BPS Branes,
hep-th/0002037.

\bibitem{DouglasFiolRomelsb} M.R.~Douglas, B.~Fiol and C.~R\"omelsberger,
The Spectrum of BPS Branes on a Noncompact Calabi--Yau,
hep-th/0003263.

\bibitem{denef} F.~Denef,
Supergravity Flows and D-Brane Stability, hep-th/0005049.

\bibitem{FM} B.~Fiol and M.~Mari\~no,
BPS States and Algebras from Quivers, 
JHEP {\bf 0007} (2000) 031, hep-th/0006189.

\bibitem{DiaconescuDouglas}
D.-E.~Diaconescu and M.R.~Douglas,
D-Branes on Stringy Calabi--Yau Manifolds,
hep-th/0006224.

\bibitem{ChKlYaZa} T.-M.~Chiang, A.~Klemm, S.-T.~Yau and E.~Zaslow,
Local Mirror Symmetry: Calculations and Interpretations,
Adv.~Theor.~Math.~Phys.~{\bf 3} (1999),
hep-th/9903053.

\bibitem{Oda}T.~Oda,
{\it Convex Bodies and Algebraic Geometry: 
An Introduction to the Theory of Toric Varieties}, 
Ergebnisse der Mathematik und ihrer Grenzgebiete,
3 Folge Band {\bf 15}, Springer-Verlag, Berlin  (1988).

\bibitem{Fulton}W.~Fulton,
{\it Introduction to Toric Varieties},
Annals of Mathematics Studies {\bf 131},
Princeton Univ. Press, Princeton (1993).

\bibitem{Ba} V.V.~Batyrev,
Variations of the Mixed Hodge Structure of Affine Hypersurfaces 
in Algebraic Tori,
Duke~Math.~J. {\bf 69} (1993) 349--409.

\bibitem{HKTY}S.~Hosono, A.~Klemm, S.~Theisen and S.-T.~Yau,
Mirror Symmetry, Mirror Map and Applications to Calabi--Yau
Hypersurfaces, Commun.~Math.~Phys.~{\bf 167} (1995) 301--350,
hep-th/9308122; Mirror Symmetry, Mirror Maps and Applications 
to Complete Intersection Calabi--Yau Spaces, 
Nucl.~Phys.~{\bf B433} (1995) 501--554, hep-th/9406055.

\bibitem{AsGrMo} P.S.~Aspinwall, B.R.~Greene and D.R.~Morrison,
Measuring Small Distances in $N\!=\!2$ Sigma Models,
Nucl.~Phys. {\bf B420} (1994) 184--242, hep-th/9311042.

\bibitem{As} P.S. Aspinwall,
Resolution of Orbifold Singularities in String Theory 
in {\it Mirror Symmetry II}, AMS/IP Studies in Adv. Math. {\bf 1},
B.R.~Greene and S.-T.~Yau (eds.),
A.M.S., Providence/International Press, Cambridge (1997) 355--379,
hep-th/9403123.

\bibitem{KlemmMayrVafa}
A.~Klemm, P.~Mayr and C.~Vafa,
BPS States of Exceptional Non-Critical Strings,
in {\it Advanced Quantum Field Theory},
J.~Fr\"ohlich {\em et al.} (eds.),
Nucl. Phys. {\bf B} (Proc. Suppl.) {\bf 58} (1997) 177--194,
hep-th/9607139. 

\bibitem{LeMayWar} W.~Lerche, P.~Mayr and N.P.~Warner,
Non-Critical Strings, del Pezzo Singularities 
and Seiberg--Witten Curves,
Nucl.~Phys. {\bf B499} (1997) 125--148,
hep-th/9612085.

\bibitem{KlemmZaslow}
A.~Klemm and E.~Zaslow,
Local Mirror Symmetry at Higher Genus,
hep-th/9906046.


\bibitem{MOY}
K.~Mohri, Y.~Ohtake and S.-K.~Yang,
Duality Between String Junctions and D-Branes on Del Pezzo Surfaces,
hep-th/0007243.


\bibitem{Bate}
{\it Higher Transcendental Functions}, Vol. I,
The Bateman Manuscript Project,
A. Erd\'elyi (ed.),
McGraw-Hill, New York (1953).

\bibitem{WhittakerWatson}
E.T.~Whittaker and G.N.~Watson,
{\it A Course of Modern Analysis}, fourth edition,
Cambridge Univ. Press, Cambridge (1927).

\bibitem{MinahanNem}
J.A.~Minahan, D.~Nemeschansky and N.P.~Warner,
Partition Functions for BPS States of the Non-Critical $E_8$ String,
Adv.~Theor.~Math.~Phys. {\bf 1} (1998) 167--183,
hep-th/9707149.

\bibitem{zeta} See, for instance, {\em From Number Theory to Physics},
M. Waldschmidt {\em et al.} (eds.), Springer-Verlag, Berlin (1995);
K.~Kato, N.~Kurokawa and T.~Saito,
{\it Number Theory 1}, Translations of Mathematical Monographs {\bf 186},
Amer.~Math.~Soc., Providence, (1999).

\bibitem{FRV}F.~Rodriguez Villegas,
Modular Mahler Measures I,
in {\em Topics in Number Theory in Honor of B. Gordon and S. Chowla},
Mathematics and its Applications {\bf 467}, 
S.D.~Ahlgrem {\em et al.} (eds.), 17--48,
Kulwer Academic Publishers, Dordrecht (1999);
also visit http:\,//www.ma.utexas.edu/users/villegas.

\bibitem{Boyd}D.W.~Boyd,
Mahler's Measure and Special Values of $L$-Functions,
Experiment. Math. {\bf 7} (1998), 37--82
; see also ftp:\,//math.ubc.ca/pub/boyd/mahler.


\bibitem{candel} P.~Candelas, X.C.~de~la~Ossa, P.S.~Green and L.~Parkes,
{A Pair of Manifolds as an Exactly Soluble Superconformal Theory},
Nucl.~Phys.~{\bf B359} (1991) 21--74.

\bibitem{OoguriOzYin}
H.~Ooguri, Y.~Oz and Z.~Yin,
D-Branes on Calabi--Yau Spaces and Their Mirrors,
Nucl. Phys. {\bf B477} (1996) 407--430, hep-th/9606112.

\bibitem{GreeneKanter}B.R.~Greene and Y.~Kanter,
Small Volumes in Compactified String Theory,
Nucl.~Phys.~{\bf B497} (1997) 127--145,
hep-th/9612181.

\bibitem{KlemmLercheMayr}
A.~Klemm, W.~Lerche and P.~Mayr,
K3-Fibrations and Heterotic-Type II String Duality,
Phys.~Lett.~{\bf B375} (1995) 313--322,
hep-th/9506112.

\bibitem{LianYau}B.H.~Lian and S.-T.~Yau,
Arithmetic Properties of Mirror Map and Quantum Coupling,
Commun.~Math.~Phys. {\bf 176} (1996) 163--191,
hep-th/9411234.

\bibitem{ConwayNorton} J.H.~Conway and S.P.~Norton,
Monstrous Moonshine,
Bull.~London Math.~Soc.~{\bf 11} (1979) 308--339.

\bibitem{Kohno}
M.~Kohno,
{\it Global Analysis in Linear Differential Equations},
Mathematics and its Applications {\bf 471},
Kulwer Academic Publishers, Dordrecht (1999). 

\bibitem{LianYau2}
B.H.~Lian and S.-T.~Yau,
Mirror Maps, Modular Relations and Hypergeometric Series I,
hep-th/9507151.

\bibitem{KannoYang}
H.~Kanno and S.-K.~Yang,
Donaldson--Witten Functions of Massless N=2 Supersymmetric
QCD,
Nucl.~Phys. {\bf B535} (1998) 512--530,
hep-th/9806015.

%
\bibitem{Deninger}C.~Deninger,
Deligne Periods of Mixed Motives, $K$-Theory and the Entropy of Certain 
${\Bbb Z}^n$-Actions,
J.~Amer.~Math.~Soc.~{\bf 10} (1997), 259--281.

\bibitem{ChYin}Y.-K.~E.~Cheung and Z.~Yin,
Anomalies, Branes and Currents,
Nucl.~Phys.~{\bf B517} (1998) 69--91,
hep-th/9710206.

\bibitem{MinaMoore} R.~Minasian and G.~Moore,
$K$ Theory and Ramond-Ramond Charge,
JHEP {\bf 9711} (1997) 002, hep-th/9710230.

\bibitem{HI}
T.~Hauer and A.~Iqbal,
Del Pezzo Surfaces and Affine 7-Brane Backgrounds,
JHEP {\bf 0001} (2000) 043,
hep-th/9910054.

\bibitem{HIV} K.~Hori, A.~Iqbal and C.~Vafa,
D-Branes and Mirror Symmetry,
hep-th/0005247.

\bibitem{Horja} R.P.~Horja,
Hypergeometric Functions and Mirror Symmetry in Toric Varieties,
math.AG/9912109.

\bibitem{hosono} S.~Hosono, Local Mirror Symmetry and Type IIA Monodromy of
Calabi--Yau Manifolds, Adv.~Theor.~Math.~Phys.~{\bf 4} (2000),
hep-th/0007071.

\bibitem{DrezetLePotier}J.-M.~Drezet and J.~Le Potier,
Fibr\'es Stables et Fibr\'es Exceptionnels sur ${\Bbb P}_2$,
Ann.~Scient.~\'Ec.~Norm.~Sup.~{\bf 18} (1985) 193--244.

\bibitem{Rudakov} A.N.~Rudakov,
The Markov Numbers and Exceptional Bundles on $\Ptwo$, 
Math.~USSR Izv.~{\bf 32} (1989) 99--112.


\end{thebibliography}
\end{document}